\documentclass[twocolumn,amsmath,amssymb,prl,10pt,nofootinbib,superscriptaddress]{revtex4}
\usepackage{ graphicx, float,amsmath, amsmath, amssymb}
\usepackage[export]{adjustbox}

\def\be{\begin{equation}}
\def\ee{\end{equation}}
\def\bea{\begin{eqnarray}}
\def\eea{\end{eqnarray}}
\def\bse{\begin{subequations}}
\def\ese{\end{subequations}}

\usepackage[breaklinks, colorlinks, citecolor=blue]{hyperref}
\usepackage[labelsep=period]{caption}
\usepackage[normalem]{ulem}
\usepackage{mathtools}
\usepackage{siunitx}
\usepackage{soul}
\usepackage{physics}
\usepackage{minitoc}

\usepackage{bm}

\DeclarePairedDelimiterXPP\BigOSI[2]%
  {\mathcal{O}}{(}{)}{}%
  {\SI{#1}{#2}}

\begin{document}
\title{The string theory swampland in the Euclid, SKA and Vera Rubin observatory era}

\author{Aurélien Barrau}%
\affiliation{%
Laboratoire de Physique Subatomique et de Cosmologie, Universit\'e Grenoble-Alpes, CNRS/IN2P3\\
53, avenue des Martyrs, 38026 Grenoble cedex, France
}


\author{Cyril Renevey}%
\affiliation{%
Laboratoire de Physique Subatomique et de Cosmologie, Universit\'e Grenoble-Alpes, CNRS/IN2P3\\
53, avenue des Martyrs, 38026 Grenoble cedex, France
}

\author{Killian Martineau}%
\affiliation{%
Laboratoire de Physique Subatomique et de Cosmologie, Universit\'e Grenoble-Alpes, CNRS/IN2P3\\
53, avenue des Martyrs, 38026 Grenoble cedex, France
}



\newcommand{\edit}[1]{{\color{red}#1}}

\date{\today}
\begin{abstract} 
This article aims at drawing the attention of astronomers on the ability of future cosmological surveys to put constraints on string theory. The fact that ``quantum gravity" might be constrained by large scale astrophysical observations is a remarkable fact that has recently concentrated a great amount of interest. In this work, we focus on future observatories and investigate their capability to put string theory, which is sometimes said to be ``unfalsifiable", under serious pressure. We show that the combined analysis of SKA, Euclid, and the Vera Rubin observatory -- together with Planck results -- could improve substantially the current limits on the relevant string swampland parameter. In particular, our analysis leads to a nearly model-independent prospective upper bound on the quintessence potential, $\abs{V'}/V<0.16$, in strong contradiction with the so-called {\it de Sitter conjecture}. Some lines of improvements for the very long run are also drawn, together with generic prospective results, underlining the efficiency of this approach. The conjectures used in this work are discussed pedagogically, together with the cosmological models chosen in the analysis. 
\end{abstract}
\maketitle

\section{Introduction}

String theory might be the only available and   viable  candidate ``theory-of-everything". It features many unique and appealing properties (see, {\it e.g.}, \cite{Polchinski:1998rq,Danielsson:2001et,Becker:2007zj,Blumenhagen:2013fgp} for pedagogical introductions). Replacing point particles by one-dimensional quantum objects leads to a completely new paradigm that has been built over the last 5 decades. Among many others, an important consequence of this fundamental shift is the natural appearance of a massless spin-2 particle, the graviton. In a sense, (quantum) gravity is therefore an unavoidable {\it prediction} of string theory. The long history of string theory  (see, {\it e.g.}, \cite{Rickles:2014fha}) went through several revolutions, from the discovery of superstrings -- and the understanding that the theory might be capable of describing all elementary particles as well as all the interactions between them -- to the unification of different versions of the model into the M-theory framework (see, {\it e.g.}, \cite{Schwarz:1996qw}). The ideas of string theory have far reaching consequences in mathematical physics, cosmology, condensed matter physics, particle physics, nuclear physics, and black hole physics. More than a well defined axiomatic model, string theory, in the broad sense, is a kind of rich and intricate framework (see, {\it e.g.}, \cite{Tong:2009np}) made of many interconnected sub-fields.\\

In spite of its extraordinary mathematical elegance -- beginning by the historical Green–Schwarz anomaly cancellation mechanism \cite{Green:1984sg} which led to the first revolution --, string theory raises questions about its falsifiability. One might provocatively argue that all the predictions made so far were somehow contradicted by observations. First, the number of dimensions required is not the one we know in nature \cite{Lovelace:1971fa}. Second, the World should be supersymmetric \cite{Gliozzi:1976qd}. Third, the cosmological constant is expected to be negative (see, {\it e.g.}, arguments in \cite{Witten:2000zk}). Fourth, non-gaussianities are to be observed in the cosmological microwave background (CMB) \cite{Lidsey:2006ia}. Of course, ways out of those na\"{\i}ve expectations are well known and it is not the purpose to describe them here in details : compactified extra-dimensions are possible, supersymmetry can be broken at a high energy scale, the introduction of membranes as new fundamental objects might stabilize a positive cosmological constant, moduli fields might not contribute simultaneously to the inflation dynamics, etc. Without going into the quantitative arguments (the interested and unfamiliar reader can have a flavor of the main ideas in, {\it e.g.}, \cite{Gubser:2010zz}) it is fair to conclude that, at this stage, string theory is not in contradiction with observations. The main concern is different. Just the other way round, one might wonder if string theory is actually falsifiable (see, {\it e.g.}, references in \cite{Smolin:2006pe,Rovelli:2011mu}). It might be feared that the intrinsic richness of the paradigm is such that basically {\it anything} happening in nature could be accounted for by string theory. The very scientific nature of string theory could then be questioned, at least in a poperrian sense. Although quite a lot of possible ``tests" were considered in the past (see, {\it e.g.}, \cite{Kane:1997fa,Casadio:1998ae,Hewett:2005iw,Kallosh:2007wm,Durrer:2011bi}), the great novelty now emerging is that cosmological surveys might severely constrain -- if not exclude -- string theory. This work aims at deriving new results on the limits that could be obtained in the next decade and at driving, pedagogically, the attention of astronomers on this somehow unexpected importance of the study of large scale structures for fundamental high energy physics.\\

As far as surveys are concerned, we focus in the following on 3 main experiments: Euclid \cite{Laureijs:2011gra,Amendola:2012ys,Amendola:2016saw}, the Vera Rubin observatory \cite{Abell:2009aa,Mandelbaum:2018ouv}, and the Square Kilometer Array (SKA) \cite{Yahya:2014yva,Maartens:2015mra,Santos:2015gra}. Euclid is a European Space Agency mission using a  Korsch-type space telescope. Its diameter is 1.2 m with a focal length of 24.5 m. The instruments are a visible-light camera (VIS) and a near-infrared camera/spectrometer (NISP). The Vera Rubin observatory, previously called LSST, is a ground based telescope whose diameter is 8.4 m with a focal length of 10.3 m. The focal plane is made of a 3.2 billion pixels matrix and the redshifts will be determined photometrically. Both are scheduled for 2022. Finally, SKA is a radio telescope array project with stations extending out to a distance of 3000 km from a concentrated central core. It will be operated in several successive stages and the fully completed array is expected around 2027. In the following, Euclid and LSST will often be mixed as they provide basically the same information in a complementary way.\\

We first explain, as intuitively as possible, the main ideas of the string theory swampland program that allow a connection between quantum gravity and very large scale -- ultra low-energy -- astronomical observations. This is rooted in the apparent impossibility to produce a real de Sitter space (positive cosmological constant) in string theory. According to the {\it de Sitter conjecture}, if we were to measure a scale factor behavior resembling too much a pure cosmological constant, the whole edifice might be in trouble. We then focus on quintessence models for dark energy and explain the specific potentials we choose together with the motivations underlying those choices to test the conjecture. The next section is devoted to the presentation of the results for each considered class. The aim is to answer this question: could the next generation of experiments show that we live in the so-called swampland, therefore excluding string theory ({\it if} the de Sitter conjecture, that will be described in details, is correct)? We conclude by underlining the limitations of the approach and possible developments to be expected for the future.

\section{The string theory swampland}

Although string theory is a tentative unification theory, it actually leads to a huge amount of false (metastable) vacua : the so-called {\it landscape} (see, {\it e.g.}, \cite{Susskind:2003kw,Banks:2003es,ArkaniHamed:2006dz,Polchinski:2006gy}). Technically, the tremendously large number of possibilities comes from choices of Calabi–Yau manifolds and of generalized magnetic fluxes over different homology cycles. More intuitively, this landscape represents a vast collection of effective low-energy theories. Otherwise stated, although the fundamental equations of string theory are often simple and elegant, their solutions are extraordinarily complex and diverse. This diversity is not necessarily a failure of physics. It may help understanding some curious features of our Universe. Just as the Earth is a very specific -- and anthropically favored -- place in the observable universe, the observable universe could be a special sample in the ``multiverse" (see, {\it e.g.}, \cite{Stoeger:2004sn,Garriga:2005av,Vilenkin:2006xv,Barrau:2007ce,Carr:2007zzb,Hall:2007ja}). In principle, the idea of a multiverse, generated for example -- but not necessarily -- by inflation, filled with different low-energy landscape realizations of string theory is {\it not} outside of the usual scientific arena: probabilistic predictions can be made and tested \cite{Freivogel:2011eg}. Having a single sample at disposal (our Universe) makes things complicated but not radically different from the usual situation in physics where the amount of information available is always finite and incomplete \cite{Carroll:2018tme}. It has never been mandatory to check {\it all} the predictions of a theory to make it scientific. Of course, providing clear predictions in a string landscape multiverse is, to say the least, challenging: from the clear definition of a measure to the accurate estimate of the anthropic weight, many questions remain open \cite{Stoeger:2004sn}. Still, the idea that string theory generates a landscape of solutions -- that can be considered as a set of effective physical theories -- remains central to the field. Although often associated with the multiverse, this wide variety of solutions can, of course, be also meaningfully considered in a single universe, which is the framework of this study.\\

The situation changed drastically with the emergence of the swampland idea \cite{Vafa:2005ui,Ooguri:2006in}. The swampland refers to the huge space of theories that seem compatible with (or possibly derived from) string theory but which, actually, are not. The landscape corresponds to the, more restricted, space of ``possibly correct" theories, in the sense that they really emerge from string theory, which is here assumed to be the framework. The swampland corresponds to theories that are not consistent when considered in details, although they appear to be correct at first sight. The landscape is a much smaller subset of the space of theories than the swampland: the island of consistent theories in a sea of incoherent proposals, even if those proposals were looking like viable candidates from a simple field theory point of view. In many works, the strategy consists in guiding the construction of effective field theories so that they belong to the landscape and not to the swampland. This is a valuable help in low energy model building. In this article, we take a different view. We try to determine the correct description of the cosmological dynamics and show that it might actually correspond to a swampland theory. If the real world were to lie in the swampland, it might strongly suggest that either string theory itself or the swampland program is wrong\footnote{we will resist the temptation to consider the possibility that string theory is correct and that it is the real world which is wrong !}. This is a unique -- although still intensively debated and clearly controversial -- way of possibly falsifying a candidate theory of quantum gravity. Surprisingly, the detailed features of the acceleration of the Universe are a privileged way to address this outstanding question.\\

Let us be slightly more specific. Excellent reviews on the swampland ideas can be found in \cite{Brennan:2017rbf,Palti:2019pca}. We simply give here a flavor of the arguments to fix the ideas of the unfamiliar reader. In the space of theories, the swampland would be vastly larger than the landscape. Theories in the swampland are apparently fine. They can be used to make prediction and look like good physics. But, when looking more carefully, they cannot be consistently completed in the ultra-violet (that is at high energy). This does not mean that they are not quantum theories of gravity but rather that they {\it cannot} appear consistently as a low energy limit of a ``string theory inspired" quantum gravity model. Otherwise stated : the coupling to gravity is problematic for those theories. One way of using this information is to discard such theories and focus on other models. However, as explained before, an interesting situation would be the one where our Universe is described by a theory belonging to the swampland:  the logic could then be reversed and the framework suggesting to discard the theory should be revised or abandoned. The swampland program is very wide and uses at the same time rigorous string theory arguments, general quantum gravitational ideas and microphysics inputs.\\

Illustratively, one can consider an effective quantum field theory (EQFT) self-consistent up to a scale $E_{self}$. If gravity is added to the game (with its own finite scale $E_{Planck}=\sqrt{\hbar c^5/G}$), the new theory will exhibit a new limit energy scale $E_{swamp}$. This is basically the energy above which the theory has to be modified if it is to become compatible with quantum gravity at very high energies. The interesting situation \cite{Palti:2019pca} is the one where $E_{swamp}<E_{self}<E_{Planck}$. Even better, if $E_{swamp}$ is smaller than any characteristic energy scale involved in the theory, it means that the whole theory lies in the swampland. This can be very helpful as a guidance in a model-building approach.

At this stage, the swampland program is mostly an ensemble of conjectures. Some of them are very reliable and -- to some extent -- demonstrated. They are theorems (even if the hypotheses are sometimes stronger than one would like). Others are speculative and grounded in extrapolations (for example from known vacua to all vacua). They rely on many different kinds of arguments, from very formal ones to qualitative {\it gedankenexperiments} involving black holes. \\

Among others, the following swampland conjectures are intensively being considered: the distance conjecture \cite{Ooguri:2006in,Klaewer:2016kiy} (it is not possible to move too much in the field space), the weak gravity conjecture \cite{ArkaniHamed:2006dz,Cheung:2014vva} (gravity is always the weakest force), the species scale conjecture \cite{Veneziano:2001ah,Dvali:2001gx} (there is a bound on the cutoff scale which depends on the number of particles), the no global symmetry conjecture \cite{Banks:1988yz,Banks:1988yz} (there cannot be an exact global symmetry in a theory with a finite number of states), the completeness conjecture \cite{Polchinski:2003bq} (if there is a gauge symmetry the theory has to incorporate states with all possible charges), the emergence conjecture \cite{Heidenreich:2017sim,Grimm:2018ohb,Ooguri:2018wrx,Harlow:2015lma} (the kinetic terms for all fields are emergent in the low-energy limit by integrating out states up to a finite scale), the non-supersymmetric AdS instability conjecture \cite{Ooguri:2016pdq} (non-supersymmetric anti-de Sitter space is unstable), the spin-2 conjecture \cite{Klaewer:2018yxi} (any theory with spin-2 massive fields has a cutoff scale\footnote{In a way, this simple fact about effective field theories significantly predates the entire Swampland program \cite{ArkaniHamed:2002sp}.}), etc. In this work we shall focus one the so-called de Sitter conjecture \cite{Obied:2018sgi,Andriot:2018wzk,Garg:2018reu,Ooguri:2018wrx}. It basically encodes the fact that it is very hard to build something resembling a positive cosmological constant in string theory. Let us describe the underlying idea.\\

The key-point is that in the 11-dimensional supergravity theory arising as a low-energy limit of M-theory, there is {\it no} cosmological constant. What is usually described as a cosmological constant actually corresponds, in this framework, to the local minimum of a scalar potential. This results from the compactification of the higher-dimensional (supersymmetric) theory. A de Sitter space does not here refer to a ``real" positive cosmological constant but, instead, to the (meta-)stable state of a scalar field with a positive value of the potential. Anything looking like a positive cosmological constant is extremely hard to construct in string theory. The profound reasons behind are, on the one hand, the instability of the so-called moduli\footnote{Moduli fields are scalar fields with properties often associated with  supersymmetric systems. They arise naturally in string theory.} in the de Sitter vacuum and, on the other hand, the non-supersymmetric nature of the de Sitter space. The heart of the conjecture lies in the bet that the tremendous difficulties arising when trying to build a de Sitter vacuum are hints that such a state does not exist at all in the theory \cite{Danielsson:2018ztv}. \\

This would not mean that the de Sitter-like behavior of the contemporary Universe -- that is its accelerated expansion (see \cite{Riess:1998cb} for the historical detection and \cite{Mortonson:2013zfa} for a brief and recent review) -- is in radical conflict with the conjecture. Just as it was probably the case for the primordial inflation (see \cite{Senatore:2016aui} for an excellent introduction), it might be that the current acceleration is obtained dynamically by a scalar field rolling down a potential. In such a case, the conjecture might hold but would constrain the shape of the potential. Although they suffer from serious drawbacks and are obviously more complicated than a pure cosmological constant \cite{Bianchi:2010uw}, such quintessence models must not appear as too strange or unnatural. Indeed, if a true cosmological constant happens to become dominant in the cosmic dynamics, it will automatically remain so forever as it does not dilute whereas the other fluids of the Universe do. It is therefore impossible that inflation was driven by the cosmological constant and the introduction of something like a scalar field evolving in an appropriate potential is required. It makes sense to assume that the contemporary phase of acceleration, which resembles the primordial one, is due to the same cause. Quintessence models for dark energy are precisely focused on this idea (see \cite{Brax:2017idh}) and receive a strong support from both particle physics and EQFTs. This might provide a solution which pleases string theory and remains compatible with what we observe. For the sake of completeness, we should however mention that quintessence is {\it not} the only alternative to a cosmological constant.\\

However, it is not sufficient to choose a dynamical dark energy model instead of a pure cosmological constant to avoid falling in the swampland. The very interesting aspects of the de Sitter conjecture is that it provides a constraint that has to be satisfied by the potential. Let us consider the usual action for gravity and scalar field $\phi$ with a potential $V$:
\begin{equation}
S=\int d^4x\sqrt{-g}\left[R-\frac{1}{2}g^{\mu\nu}\partial_{\mu}\phi\partial_{\nu}\phi-V(\phi)\right],
\end{equation}
where $R$ is the Ricci scalar\footnote{The forthcoming statements can be generalized to a set of scalar fields, introducing a supermetric $G_{ij}$ in the field space, but we prefer to keep the notations simple at this stage.}. Unless otherwise stated we use Planck units throughout all this work. The original de Sitter conjecture \cite{Obied:2018sgi} states that 
\begin{equation}
\frac{|V'|}{V}>\lambda_c,
\end{equation}
where $\lambda_c$ is a constant of order one, and the derivative is to be understood with respect to  the field. Interestingly, recent studies set a clear numerical numerical value in the large field limit \cite{Bedroya:2019snp,Andriot:2020lea}: $\lambda_c>\sqrt{2/3}\approx0.82$. It has already been shown \cite{Agrawal:2018own} that current observations lead to $|V'|/V<0.6$, which is in tension with the conjecture. We will show that the Euclid/Rubin/SKA generation of experiments should lead to a clear improvement of this limit and potentially show that the actual behavior of the Universe unambiguously does not satisfy the conjecture.\\

This simple version of the de Sitter conjecture is however not sufficient. The well-understood Standard Model of particle physics allows one to extrapolate the existence of other critical points of the potential. This argument was first made for the SM Higgs potential \cite{Denef:2018etk} where some loopholes were pointed out in the argument. It was then elaborated on in \cite{Murayama:2018lie} and extended in \cite{Choi:2018rze} to consider the pion potential leading to the firm conclusion that there is no way that a quintessence model coupled to the Standard Model can be consistent with the original de Sitter Swampland conjecture that bounds only $|V'|/V$. Avoiding a critical point of the pion potential would require a large quintessence coupling in violation of stringent fifth-force constraints. These arguments led to a refined form of the bound so that the potential has tachyonic directions in the regions where $V'$ is too small\footnote{One should remain careful to avoid a possible loophole in the reasoning. If the conjecture is modified each time a model describing correctly our World contradicts it, there is no way it can be used to falsify the framework in which the conjecture is established. It is therefore mandatory to strengthen, in the near future, the theoretical ground on which the de Sitter conjecture is derived.}.

The refined conjecture reads \cite{Ooguri:2018wrx,Garg:2018reu}:
\begin{align}
    \frac{\abs{V'}}{V}>\lambda_c\quad\textrm{or}\quad
    \frac{V''}{V}<-\alpha_c, \label{swamp_refined}
\end{align}
where $\lambda_c$ and $\alpha_C$ are both positive numbers of order one. In addition to the previously given arguments, the second condition also comes from the fact that when it is satisfied, an instability develops, leading to a beakdown of the entropy-based argument for the first condition. In the following, we first consider the original conjecture and then comment on the refined version. It should also be mentionner that several other improvements are also being considered \cite{Garg:2018reu,Andriot:2018wzk,Ben-Dayan:2018mhe,Dvali:2018jhn,Garg:2018zdg,Andriot:2018mav}. \\

Why are we to expect the derivative of the potential (normalized to the potential) not to be too small in string theory ? The reasons are subtle \cite{Ooguri:2018wrx,Garg:2018reu} and no obvious argument can be given. De sitter space is a highly non-trivial structure. It has a horizon and  is endowed, analogously to what happens for black holes, with a temperature and an entropy. This is expected not to be stable in quantum gravity. The de Sitter conjecture can be seen as resulting from the distance conjecture \cite{Ooguri:2018wrx,Geng:2019zsx}. Very loosely speaking this is because the distance traveled in the field space is linked with the dimension of the Hilbert space of the theory which is itself (potentially) linked with the entropy of the considered de Sitter space.  It can also be approached from thermodynamical considerations \cite{Seo:2019mfk}: under several assumptions, it might be that when the number of degrees of freedom is enhanced as the modulus rolls down the potential, the bound on $|V'|/V$ becomes equivalent to the condition for the positive temperature phase. As mentioned before, the original de Sitter conjecture has been refined \cite{Andriot:2018wzk,Andriot:2018ept,Andriot:2018mav} and there are now quite a lot of arguments supporting its validity although the accurate value of the bound $\lambda_c$ is still debated. Let us introduce $\lambda\equiv -V'/V $ (which depends on $\phi$, hence on time) and summarize: if a potential resembling too much a cosmological constant (that is $|\lambda|\ll 1$) were to be measured, it would hardly be compatible with string theory if the conjecture is correct. \\

In principle, studying the primordial inflation leads to more stringent constraints on $\lambda$ than considering the contemporary acceleration of the Universe. Although inflation is unquestionably part of the cosmological paradigm, it is however fair to underline that other scenarii are also being considered: strictly speaking inflation is not proven. Several alternatives are described in \cite{Durrer:1995mz,Hollands:2002yb,Veneziano:2003enc,Brandenberger:2009jq,Creminelli:2010ba,Poplawski:2010kb,WilsonEwing:2012pu,Lilley:2015ksa} and references therein. If a study aims at investigating how the actual behavior of the Universe constrains -- or even might exclude -- string theory, it is mandatory to rely on established facts. For this reason, we focus on the late time acceleration that is being directly observed. 

\section{Dark energy potentials}

\subsection{General picture}

The acceleration of the Universe is now unquestionable. This has been observed  using type Ia supernovae \cite{Perlmutter:1998np}, using the cosmological microwave background (CMB) \cite{Aghanim:2018eyx}, and using baryon acoustic oscillations (BAO) \cite{Blake:2011en}.
In a way, a pure cosmological constant -- as a part of the full Einstein's equations according to Lovelock's theorem -- works well to explain this acceleration \cite{Bianchi:2010uw}. If the quantum vacuum fluctuations are taken into account the question becomes merely the one of a suitable renormalization. This, however, raises many questions (see, {\it e.g.}, \cite{Brax:2017idh}) and the idea of quintessence is intensively studied to overcome most coincidence problems (see \cite{Zlatev:1998tr,Martin:2008qp,Tsujikawa:2013fta} for reviews).
Any model of quintessence must not only produce a scalar field potential that is sufficiently flat, but also fine-tune away the cosmological constant.  From the point of view of our approach it is important to consider such models as a pure cosmological constant lies trivially in the swampland if the de Sitter conjecture holds. To be conservative on a possible ``exclusion" of string theory, it is necessary to focus on dynamical models. \\

It is not possible to derive limits on $\lambda$ that are strictly model-independent. Exactly as it is not possible to obtain observational limits on the equation of state (EOS) of dark energy as a function of the redshift without assuming a given parametrization. One needs to consider a specific quintessence model, that is a potential on which the scalar field is rolling, and compare the cosmological trajectories with current -- or future -- data to derive limits on $\lambda$. We shall discuss the precise methodology in the next section but it is worth emphasizing now the deep limitation of our approach. We will consider different classes of quintessence models that are the main paths discussed and conservatively assume that the bound to keep for the analysis is the less stringent one. It is, however, not possible to exclude that a future model might relax our limits. To minimize this risk we investigate very different potentials corresponding to extremely different philosophies.\\

The second Friedmann equation, obtained by combining the first one with the trace of the Einstein's equations reads, without a cosmological constant and for a flat space:
\begin{equation}
\frac{\ddot{a}}{a}=-\frac{1}{6}\sum_i\left( \rho_i+3p_i\right),
\end{equation}
where $a$ is the scale factor and $\rho_i$ and $p_i$ are, respectively, the energy density and pressure for the fluid of type $i$. In a universe filled with matter ($p_m=0$), radiation ($p_R/\rho_R=1/3$), and a scalar field $\phi$ (with energy density $\rho_{\phi}$ and pressure $p_{\phi}$), this means
\begin{equation}
\frac{\ddot{a}}{a}=-\frac{1}{6}\left( \rho_m+2\rho_R+\rho_{\phi}+3p_{\phi}\right).
\label{fr2}
\end{equation}
A dark energy quintessence model must therefore ensure that the scalar field dominates over the other components and exhibits a negative enough pressure to achieve $\ddot{a}>0$. in Eq. (\ref{fr2}), the field pressure $p_{\phi}=\frac{1}{2}\dot{\phi}^2-V$, is the only term which can be negative. In practice, this is realized by the domination of the potential energy over the kinetic energy of the field, hence the usual reference to flat enough potentials. The tricky part lies in the fact that this domination of the negative pressure of the scalar field over all the other components has to happen very late in the cosmological history. And, if possible, for a wide range of initial conditions. This set of constraints has basically led to 3 classes of relevant potentials that we have used to study the de Sitter conjecture in this work. For all of them, we define the EOS parameter $w\equiv p_{\phi}/\rho_{\phi}$. Intuitively speaking, in freezing models the motion of the field gradually slows down because the potential becomes flat at low redshift. In thawing model, the field was initially frozen due to the Hubble friction (as during inflation) and it started evolving when the Hubble rate became small enough. Thawing models have a value of $w$ which began near -1 and increased with time, whereas freezing models have a value of $w$ which decreased, usually (but not necessarily) toward -1 \cite{Scherrer:2007pu,Chongchitnan:2007eb,Duary:2019dfu}. In a way, freezing models are more obviously answering the question of producing the required behavior of the Universe but thawing models are easier to build and, in this sense, more natural. 

Using the Friedmann and Klein-Gordon (K-G) equations, one can write first order differential equations for the parameters $w$, $\Omega_\phi=\rho_\phi/(3H^2)$ and $\lambda$, with respect to the number of e-folds $N$:
\begin{align}
\dv{w}{N}&=(w-1)\left[3(1+w)-\lambda\sqrt{3(1+w)\Omega_\phi}\right],\label{eq:w}\\
\dv{\Omega_\phi}{N}&=-3w\Omega_\phi(1-\Omega_\phi),\label{eq:Op}\\
\dv{\lambda}{N}&=-\sqrt{3(1+w)\Omega_\phi}(\Gamma-1)\lambda^2,\label{eq:l}
\end{align}
where $\Gamma \equiv VV''/(V')^2$. Rewriting the equations of motion under the form \eqref{eq:w}-\eqref{eq:l} is very useful as we are ultimately interested in constraining $\abs{\lambda}$ using constraints on $w_a\equiv -\dd w/\dd N|_{N=0}$ and $w_0\equiv w|_{N=0}$ (with $a_0=1$). Furthermore, as observations show that $\Omega_{\phi, 0}\sim 0.7$ \cite{Aghanim:2018eyx}, where the subscript $0$ refers to contemporary values, we consider that physical trajectories have to fulfill $\Omega_{\phi,\textrm{past}}\ll 1$ and $\Omega_{\phi, 0}\sim 0.7$, which is easily implementable with Eqs. \eqref{eq:w}-\eqref{eq:l}. In addition, these differential equations describe directly the parameters in which we are interested, as opposed to the Friedmann and K-G equations, thus improving the intuitive interpretations.

\subsection{Tracking freezing models}

A very effective class of potentials fulfilling the previous conditions leading to {\it tracking freezing} solutions are the Ratra-Peebles potentials \cite{Peebles:1987ek,Ratra:1987rm}:
\begin{equation}
V(\phi)=M^{4+\alpha}\phi^{-\alpha},\label{eq:potential_tracking}
\end{equation}
where $M$ is a characteristic energy scale and $\alpha >0$.
The remarkable feature of those potentials lies in the fact that the scalar field ``tracks" the background evolution. This means that the EOS parameter of the field, $w$, changes at the transition between radiation domination and matter domination. The field adapts itself to the scale factor behavior. The density of the field can be shown to decrease less rapidly than the one of the surroundings, ensuring the late time domination. \\

In addition of being a tracker, this solution is also an attractor. Impressively, the attractor solution is joined before present times for more than 100 orders of magnitude in the initial energy density of the field \cite{Martin:2008qp}. If $\alpha$ is large enough, it is possible that the energy scale $M$ of the theory becomes high whereas the energy scale of the acceleration remains very small, which has far reaching consequences for naturalness. This is somehow reminiscent of the so-called see-saw mechanism in particle physics. General considerations on tracking solutions are given in \cite{Steinhardt:1999nw} where it is, in particular, shown that the very existence of a tracker depends on the behavior of the function $\Gamma(\phi)$, which needs to be greater than one at any time. In the case of the potential described by Eq. \eqref{eq:potential_tracking}, $\Gamma=1+1/\alpha$ and the tracker condition is always satisfied.\\

\begin{figure}
    \centering
    \includegraphics[width=0.8\linewidth]{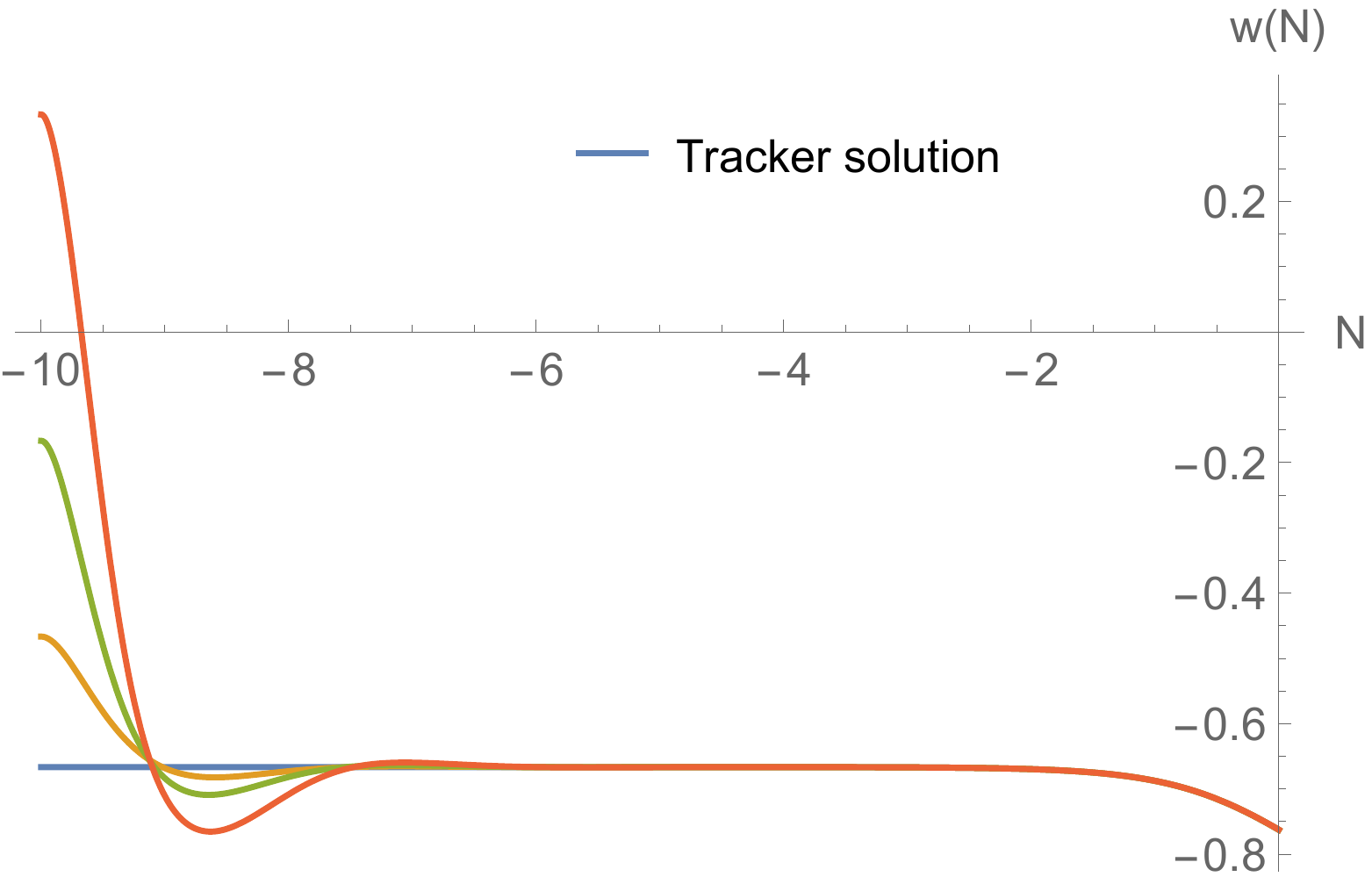}
    \caption{Stability of the solution for {\it tracking freezing} models. The blue curve represents the tracker trajectory, while the yellow, green and orange curves represent trajectories starting with various initial conditions for $w(N_i)$. }
    \label{fig:stability_tracker}
\end{figure}

The dynamics makes the system nearly insensitive to initial conditions. Whatever the reasonable choices made in the remote past, at the initial time $N_i$, trajectories converge to the tracker solution before $N=0$. One aspect of this behavior is illustrated in Fig.~\ref{fig:stability_tracker}. In addition, since the parameter $\lambda$ is directly related to $w$ and $\dd w/\dd N$ through Eq. \eqref{eq:w}, the stability of the EOS parameter under changes in the initial conditions is reflected in the behavior of $\lambda$. It can be shown, using Eqs. \eqref{eq:w}-\eqref{eq:l}, that when the EOS parameter $w$ is quasi-constant -- that is in the tracker regime -- one has: 
\begin{align}
    w\approx w_i:=\frac{-2}{2+\alpha}\quad\textrm{and}\quad
    \lambda^2\approx \lambda_i^2:=3\frac{1+w}{\Omega_\phi}. \label{eq:IC_tracking}
\end{align}
When adding the requirement that $\Omega_{\phi,0}\approx 0.7$, the system is fully determined.\\

As a side effect, this raises a quite interesting epistemological question. In the limit $t\to \infty$, this potential leads to $w\to -1$ and $-V'/V\rightarrow 0$. This will inevitably violate the de Sitter conjecture at some point in the future. Does it, however, makes sense to use a cosmological behavior that has not yet taken place to rule out a model ? Or, at least, should we use the expected future dynamics in our sets of constraints ? Although the answer is not obvious and deserves to be debated, we will remain conservative and resist the temptation to extrapolate beyond the contemporary epoch. First, because, strictly speaking, the future is not written. With the cosmic fluids considered so far, the evolution is obviously known but nothing prevents surprises (other fields, exotic matter, etc.). Second, because de Sitter space might be unstable and ``decay" before the conjecture is violated. Third, because it is a kind of ``safe" principle to assume that physics is to describe what {\it does} exist and not what might or will exist.

\subsection{Scaling freezing models}

The so-called {\it scaling freezing} models typically rely on potentials of the form \cite{Ferreira:1997hj}
\begin{equation}
V(\phi)=V_0e^{-\lambda\phi},\label{eq:potential_single_exponential}
\end{equation}
where $V_0$ and $\lambda$ are constants. The scaling solution itself corresponds to $\Omega_m/\Omega_{\phi}=constant$, where the $\Omega_k$ are the densities normalized to the critical density. It was shown \cite{Copeland:1997et} to be realized at the fixed point:
\begin{eqnarray}
\frac{\dot{\phi}}{\sqrt{6}H}&=&\sqrt{\frac{3}{2}}\frac{1+w_m}{\lambda},\\
\frac{\sqrt{V}}{\sqrt{3}H}&=&\sqrt{\frac{3(1-w_m^2)}{2\lambda^2}}.
\end{eqnarray}
Such a potential also exhibits a tracking-like solution that acts as an attractor for physical trajectories with different initial conditions. In this case, the function $-V'/V$ is constant and can be seen as a simple parameter. The system is solely described by Eqs. \eqref{eq:w} and \eqref{eq:Op}. Moreover, the stable tracking solution has a quasi-constant EOS parameter $w\approx w_i \approx -1$ and, as $\Omega_\phi$ is initially set to have $\Omega_{\phi, 0}\sim 0.7$, the system is fully determined. 

More refined models were constructed \cite{Sahni:1999qe,Albrecht:1999rm}, involving two exponential functions 
\begin{equation}
V(\phi)=V_1e^{-\lambda_1\phi}+V_2e^{-\lambda_2\phi},\label{eq:potential_double_exponential}
\end{equation}
where $V_i$ and $\lambda_i$ are constants. Constraints on the parameters have been derived \cite{Chiba:2012cb} so as the model to be consistent with CMB and BAO data. Contrary to the previous potentials, neither $\lambda$ nor $\Gamma$ is independent of $\phi$ and Eqs. \eqref{eq:w}-\eqref{eq:l} are ill-defined. Fortunately, one can easily show that $\lambda(\phi)$ is bijective which allows one to consider $\Gamma$ as a function of $\lambda$ without affecting the system. Once again, the attractor nature of the tracking solution also allows to properly define the set of initial conditions. In this case, the EOS parameter $w$ is quasi-constant for $w\approx w_i \approx 0\implies \lambda\approx \lambda_i:=\sqrt{3/\Omega_\phi(N_i)}$ with $\Omega_\phi(N_i)$ being set so that that $\Omega_{\phi, 0}\sim 0.7$.


\subsection{Thawing models}

In {\it thawing models}, the field is initially frozen with $w\approx -1$ and then begins to evolve. This is typically produced by potentials like \cite{Frieman:1995pm}
\begin{align}
V(\phi)&=V_0\left(1+\cos(\sqrt{2}\,\phi/f)\right)\quad \textrm{or}\label{eq:potential_1pcos}\\
V(\phi)&=V_0\cos(\phi/f),\label{eq:potential_cos}
\end{align}
where $V_0$ and $f$ are constants. Although the phenomenologies associated with those potentials are very close, we keep both expressions as some subtleties might differ. In this case, the evolution begins at late times but has to remain weak to account for data. Constraints were derived in \cite{Dutta:2008qn,Gupta:2011ip}. The behavior of the thawing dynamics around the redshift interval $0.6<z<1$ is quite sensitive to deviations from a pure cosmological constant \cite{Sen:2009yh}. The potentials we choose here are not the only possible ones but allow to catch most of the dynamics for this kind of models. \\

To study the behavior of $w(N)$ with the potentials given by Eqs. \eqref{eq:potential_1pcos} and \eqref{eq:potential_cos}, we use an analytical approximation 
\cite{Dutta:2008qn,Scherrer:2007pu} which reads:
\begin{align}
    w(a)\approx-1+(1+w_0)a^{3(K-1)}\mathcal{F}^2(a), \label{eq:approx_w_thawing}
\end{align}
where 
\begin{align}
   &\mathcal{F}(a)\equiv \nonumber\\ &\frac{(K-F(a))(F(a)+1)^K+(K+F(a))(F(a)-1)^K}{(K-\Omega_{\phi, 0}^{-1/2})(\Omega_{\phi, 0}^{-1/2}+1)^K+(K+\Omega_{\phi, 0}^{-1/2})(\Omega_{\phi, 0}^{-1/2}-1)^K},
\end{align}
with
\begin{align}
    K&\equiv \sqrt{1+\frac{4}{3}c_i^2}, \\ F(a)&\equiv \sqrt{1+(\Omega_{\phi, 0}^{-1}-1)a^{-3}}.
\end{align}
In addition, one introduces $c_i^2=-V''(\phi_i)/V(\phi_i)$. The approximation is valid when $w\approx -1$ and the scalar field is close to the top of its potential, which is suitable for the considered  case. It is then easy to show that $c^2=-\Gamma \lambda^2$. In the slow-roll regime, $\lambda(\phi)$ is bijective and can be inverted to get $\Gamma(\lambda)$, leading to a useful relation between $c^2$ and $\lambda$ for the potentials of Eqs. \eqref{eq:potential_1pcos} and \eqref{eq:potential_cos}:
\begin{align}
    c^2&=\frac{-\lambda^2}{2}+\frac{1}{f^2},\\
    c^2&=\frac{1}{f^2},
\end{align}
respectively.\\

In the previously considered models, the parameter $|\lambda(N)|$ was found to be decreasing (or constant) with $N$ through Eq. \eqref{eq:l}. Using the contemporary constraints on the swampland conjecture was therefore the most efficient way to go. This is not the case for thawing models where $|\dd\lambda/\dd N| > 0$ requiring to take into account the past evolution of $\lambda(N)$. To obtain Eq. \eqref{eq:approx_w_thawing}, the de Sitter solution for $\Omega_\phi(N)$ was used, namely
\begin{align}
    \Omega_\phi(a)=\frac{\Omega_{\phi, 0}\,a^3}{\Omega_{\phi, 0}\,a^3+1-\Omega_{\phi, 0}}.\label{eq:de-sitter_omega}
\end{align}
Approximations given by Eq. \eqref{eq:approx_w_thawing} and Eq. \eqref{eq:de-sitter_omega}, together with the ordinary differential equation (ODE) \eqref{eq:l} give the past evolution of $-V'/V$.

\section{Results}
 
Links between the swampland ideas and astronomical observations have already been studied in \cite{Agrawal:2018own,Raveri:2018ddi,Akrami:2018ylq,Arjona:2020skf}. In this work, however, we do not focus on existing data or question the validity of the de Sitter conjecture but, instead, we try to probe its exclusion power from the viewpoint of future surveys, as initiated in \cite{Heisenberg:2018yae}. 

\subsection{Experimental projections}

We consider on the one hand the Vera Rubin (formerly called LSST) observatory and the Euclid satellite that we shall refer to as Large Optical Surveys (LOSs) and, on the other hand, the SKA radioastronomy interferometric project. The aim is to investigate the limits they shall put on $|V'|/V$ and if the dramatic improvement in sensitivity might establish that we actually live in the Swampland (under the assumption that the de Sitter conjecture is correct), therefore suggesting that the string-inspired ideas behind this concept are incorrect.\\

Establishing in details the sensitivity of future surveys to a given cosmological observable is tricky. In the specific case we are interested in, it is tempting to use as much information as possible and, in particular, to take into account the observational constraints obtained on the equation of state parameter as a function of the redshift, $w(z)$, so as to compare the resulting curve with the expectations calculated for a given dark energy potential. This is the strategy followed in \cite{Zlatev:1998tr}. This however cannot be straightforwardly extrapolated to future experiments for which such detailed investigations are not yet available and, more importantly, this anyway relies on an assumed parametrization for the evolution of the equation of state parameter. For those reasons, we prefer to, conservatively, use only the information on $w_0$ and $w_a$, assuming the usual form $w(a)=w_0 +(1-a)w_a$ \cite{Chevallier:2000qy}, where $a$ is the scale factor. We have checked that the results obtained using exclusively $w_0$ and $w_a$, with an the exponential potential, are extremely close to those derived in \cite{Zlatev:1998tr}: the sensitivity remains practically the same.\\

The main difficulty when focusing on forecasts is to evaluate correctly the theoretical uncertainties on non-linear scales which become particularly relevant for the next generation of instruments that will probe the details of the growth of large scale structures with an extraordinary precision, up to redshifts of order 3. Billions of galaxies will be precisely located by the LOSs. The reionization era and the cosmic dawn, up to $z =20$, are even expected to be probed by SKA. The accurate evaluation of the constraints put on dark energy crucially depends on the way small scales are taken into account. This is a highly complex problem which involves general relativistic corrections to the structure formation mechanisms \cite{Tansella:2017rpi}, the galaxy nonlinear bias \cite{Jennings:2015lea}, the intrinsic alignment problem \cite{Hilbert:2016ylf}, the  feedback of baryons \cite{Schneider:2015wta}, etc. The usual strategy has been to implement a cutoff scale below which data cannot be used. This obviously misses quite a lot of potentially relevant information and it has been shown that non-linear scales could be used in a controlled way \cite{Baldauf:2016sjb}. The key-point lies in a correct estimation of the evolution of theoretical uncertainties at very high wavenumbers.

\begin{figure}
    \centering
    \includegraphics[width=0.8\linewidth]{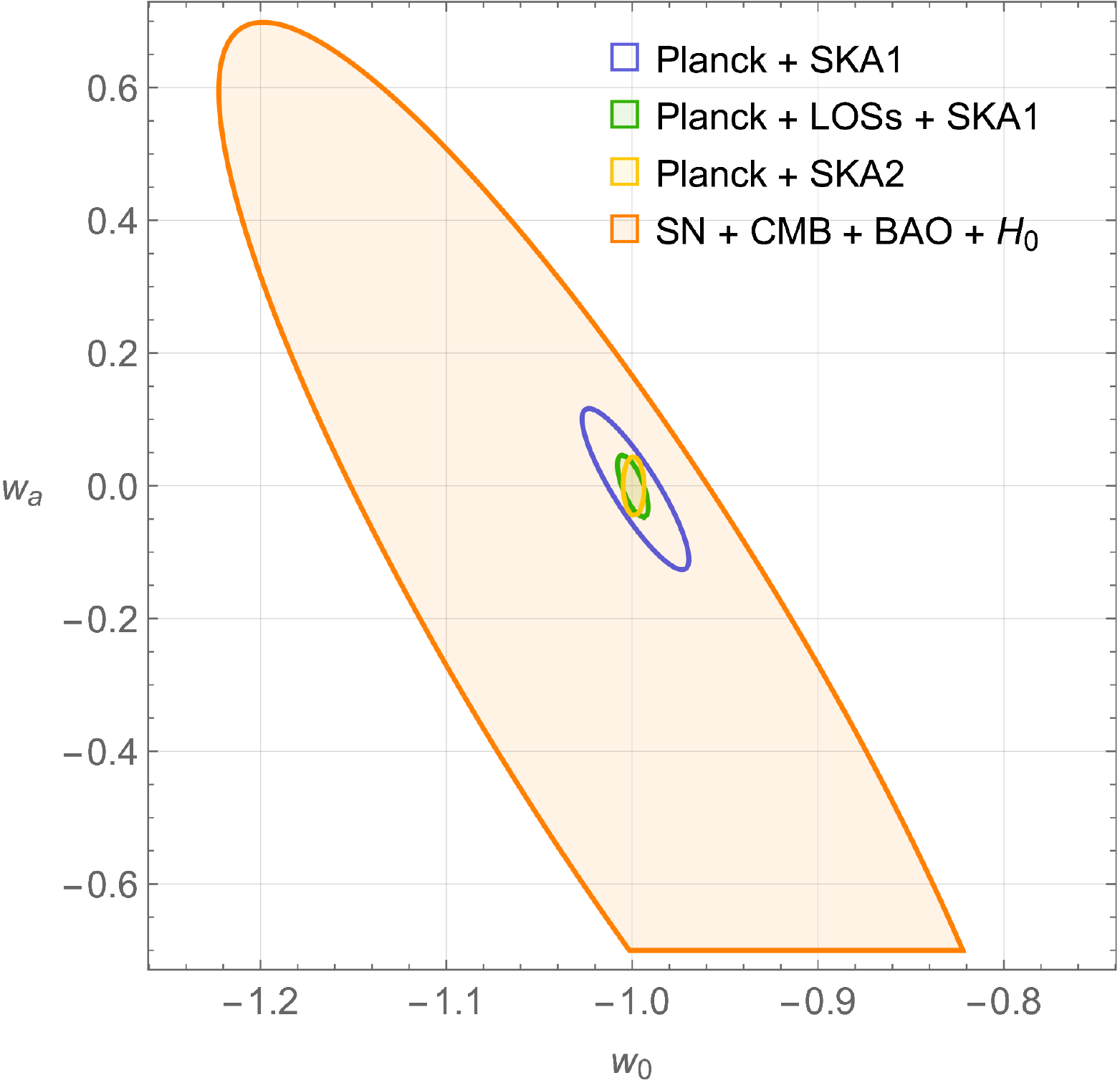}
    \caption{Comparison of the 95\% CL constraints on $w_0$-$w_a$: current results (in orange) and expected improvements (in blue, green and yellow).}
    \label{fig:compare_data}
\end{figure}

Instead of crude assumptions relying on either fully correlated or totally uncorrelated errors at small scales, a realistic numerical method to estimate uncertainties on the nonlinear spectrum was developed in  \cite{Sprenger:2018tdb}, following ideas of \cite{Audren:2012vy}. A reasonable improvement of the modelisation of nonlinear effects is  obtained by taking into account the increase of numerical resources expected by the time the real data will be available. To this aim, a Bayesian Markov–Chains–Monte–Carlo (MCMC) was used instead of the Fisher matrix formalism, subject to numerical instabilities. As massive neutrinos play an important role in the non-linear growth of structures, in addition to $w_0$ and $w_a$, the total neutrino mass $M_{\nu}$ is also left as a free parameter. It is marginalized over in the $(w_0,w_a)$ contour that we use. We basically consider 3 cases extracted from \cite{Sprenger:2018tdb}: scenario 1 is ``Planck+SKA1", scenario 2 is ``Planck+LOSs+SKA1", scenario 3 is ``Planck+SKA2". The details of the SKA1 and SKA2 programs are given in \cite{Sprenger:2018tdb,Bull:2015lja,Harrison:2016stv,Bonaldi:2016lbd}. Three  probes are included: galaxy clustering, weak lensing (for LOSs and SKA), and H{\MakeUppercase{\romannumeral 1}} intensity mapping (for SKA) at low redshift -- probing the reionized universe. We draw on Fig.~\ref{fig:compare_data} the comparison between current constraints on $w_0$ and $w_a$ \cite{Scolnic:2017caz} and what is to be expected in the future. Throughout all this study, we approximate the constraints by ellipses that match closely the numerical results. \\

Although the actual path might slightly differ in some cases (as it will be detailed below), the methodology is as follows. To evaluate the ``exclusion power" of future surveys, we assume that the actual cosmological behavior will be de Sitter-like and investigate how this might contradict the relevant swampland conjecture. For a given potential family, we vary both the initial conditions (leading to trajectories compatible with the known features of the Universe) and the values of the parameters entering the model. For each simulation, we evaluate $\abs{\lambda}=\abs{V'}/V$ along the trajectory and keep its most relevant (that is smallest) value. Then, to remain conservative, we keep the highest of those $\abs{\lambda}$ values within a given confidence level (CL) ellipse in the $w_0-w_a$ plane for different forecasts. This sets the result given in the corresponding table. To summarize intuitively: for each parameter choice and initial conditions, we compute $(w_0,w_a)$ and $-V'/V$ and evaluate how observational constraints on the first can constrain the second, in the most reliable and conservative way. In practice, the effect of the initial conditions -- provided that the model accounts for the current observations -- can hardly be noticed.

\subsection{Model-independent analysis and single-exponential potential}

\begin{figure}
    \centering
    \includegraphics[width=0.8\linewidth]{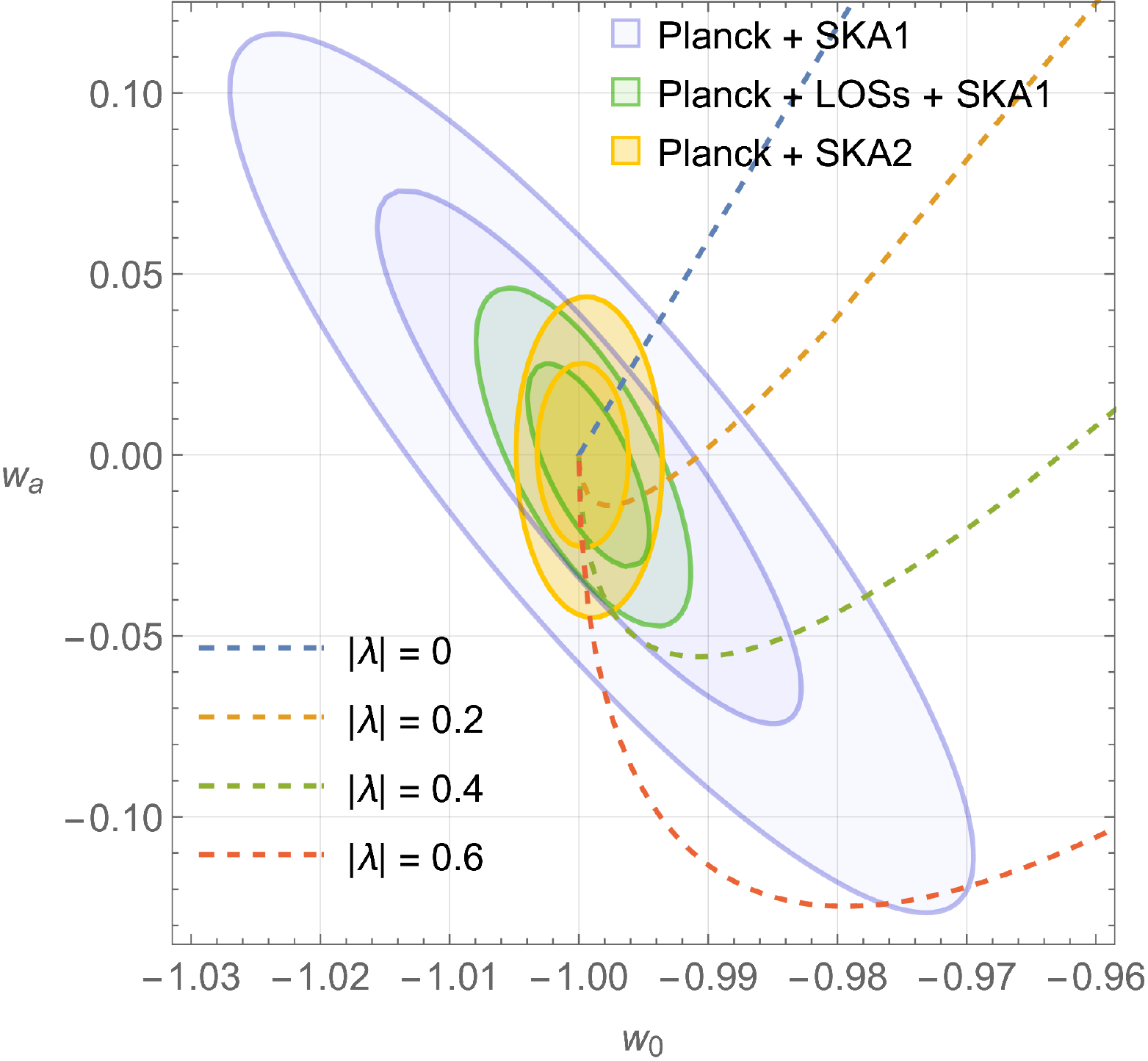}
    \caption{Relation between $w_0$, $w_a$ and $\abs{\lambda_0}=\abs{V'(0)}/V(0)$. The ellipses represent the expected 67 and 95\% CL constraints on $w_0$-$w_a$ from various future experiments, while the dashed lines are the isolines $\lambda=constant$ as calculated with equation \eqref{eq:l} evaluated at $N=0$. }
    \label{fig:lambda_independent}
\end{figure}

From the differential equation \eqref{eq:w} evaluated at $N=0$, we can directly relate $w_0$, $w_a$ and $\lambda_0:=\lambda(0)$ as measured today. We draw on Fig.~\ref{fig:lambda_independent} the isolines $\lambda=constant$ against the future constraints from Plank, Euclid and SKA, with $\Omega_{\phi, 0}\approx0.7$. As a first observation, one can see that for higher values of $\lambda$ the isoline is getting closer to the line $w_0=-1$ with $w_a\in \mathbb{R}^-$. In the limit $\lambda\rightarrow \infty$, one recovers the vertical axis. This seemingly constitutes a drawback for constraining $\abs{V'}/V$ today using a $w_0$-$w_a$ analysis. However, this conclusion fails to capture the dynamics of the system. In practice,  models leading to $w_0\approx -1$ and $w_a<0$ suffer from a major conceptual problem. Since the EOS parameter is bounded from below, $w(N)>-1$ (we do not consider here exotic phantom models known to exhibit an unstable vacuum filled with negative mass particles), such a configuration is highly unstable, which raises strong coincidence and fine-tuning problems. It is however not possible to define a model-independent constraint on $\abs{\lambda}$ from this straightforward analysis and one needs to study different potentials to test the swampland conjecture. This is the aim of te next sections. Nonetheless, it is possible to understand from Fig.~\ref{fig:lambda_independent} that producing a theoretical prediction for $w_0$ and $w_a$ for a specific model leads to a direct constraint on the parameter $\lambda_0$.\\

\begin{figure}
    \centering
    \includegraphics[width=0.8\linewidth]{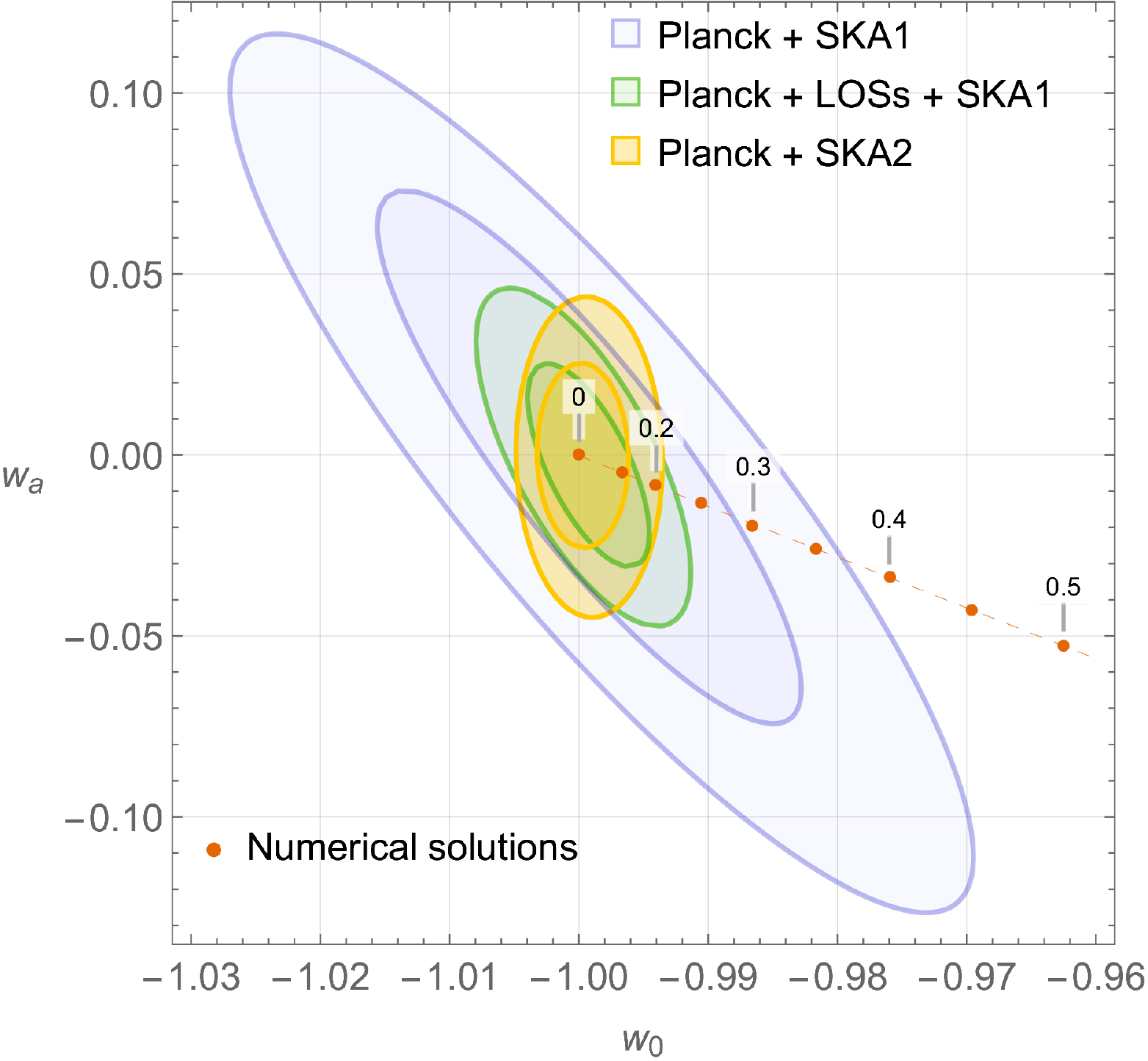}
    \caption{Comparison between the expected 67 and 95\% CL constraints on $w_0$-$w_a$ for various future experiments (blue, green and yellow ellipses) together with numerical results for $\abs{\lambda}$ for an exponential potential, as given by Eq. \eqref{eq:potential_single_exponential} (orange dots). The parameter $\lambda$ varies from $0$ to $0.5$.}
    \label{fig:constant_lambda}
\end{figure}

To better understand the method used to test the de Sitter conjecture, we start with the simplest model based on a single-exponential potential given by Eq. \eqref{eq:potential_single_exponential}. As mentioned earlier, the initial conditions are set to follow the tracker solution and to get $\Omega_{\phi, 0}\approx 0.7$. Numerical simulations with the parameter $\lambda$ ranging from $0$ to $0.5$ are displayed in Fig.~\ref{fig:constant_lambda}. Simulations for higher values of $\lambda$ were also carried out to ensure that the results follow the same trend. It is not useful to scan the parameter space for arbitrary $\lambda\in \mathbb{R}^+$ as for $\lambda\gtrsim 2$ one cannot reach $\Omega_{\phi,0}\approx 0.7$. It can indeed be shown \cite{Tsujikawa:2013fta} that for $\lambda^2>3$, the fixed point is $w=0$ and for $\lambda\gtrsim 2$ it is reached before $\Omega_\phi$ approaches $0.7$. Interestingly, the numerical results for the current values of $w$ and $-\dd w/\dd a$, for different values for $\lambda$, are aligned on a straight line (at least up to the point for $\lambda=2$). The intersections between this line and the ellipses of Fig.~\ref{fig:constant_lambda} representing the expected 67\% and 95\% confidence level of three different future experiments lead to constraints on $\abs{\lambda}=\abs{V'}/V$. The results are summarized in Table \ref{tab:single_exponential}. Remarkably, the higher upper bound obtained ($\abs{\lambda}<0.16$ at 67\% CL when using all the cosmological information that will be available) is significantly better than the limit currently available ($\abs{\lambda}<0.6$) and much smaller that the lower bound suggested by the de Sitter conjecture ($\abs{\lambda}>0.8$). \\


\begin{table}
\renewcommand{\arraystretch}{1.3}
\begin{tabular}{|c||c|c|c|}
\hline
     & Pl. + SKA1 & Pl. + LOSs + SKA1 & Pl. + SKA2 \\ \hline\hline
67\% CL & $\abs*{\lambda}<0.28$    & $\abs*{\lambda}<0.17$             & $\abs*{\lambda}<0.16$    \\ \hline
95\% CL & $\abs*{\lambda}<0.36$    & $\abs*{\lambda}<0.22$             & $\abs*{\lambda}<0.20$    \\ \hline
\end{tabular}
\caption{Expected constraints on $\abs*{V'}/V$ for an exponential potential, as given by Eq.  \eqref{eq:potential_single_exponential}. Those constraints are also valid for generic {\it scaling freezing} models described by Eq. \eqref{eq:potential_double_exponential}.}
\label{tab:single_exponential}
\renewcommand{\arraystretch}{1}
\end{table}

Non-trivially, as we shall notice for each model, and as we have just mentioned it, numerical results are ``aligned". One can then calculate the constraint on $\abs{\lambda_0}$ (which is, for the potentials considered so far, the most stringent case) by evaluating the intersection between the line and the ellipses. This is actually not obvious from the shape of the isolines $\lambda_0=constant$ in Fig~\ref{fig:lambda_independent}. One could indeed have expected that if the line of numerical results was exhibiting a strong positive slope, the intersection between the ellipses and the line would not be the relevant constraint on $\lambda_0$, but would instead be weaker. However, one can show that this is not the case, as the slope would have to be greater than 6, which is actually the slope of the isoline $\lambda_0=0$. Furthermore, one can notice that the lower the slope of the line of numerical results, the weaker the constraint on $\abs{V'}/V$.

\subsection{Tracking freezing models}
\begin{figure}
    \centering
    \includegraphics[width=0.8\linewidth]{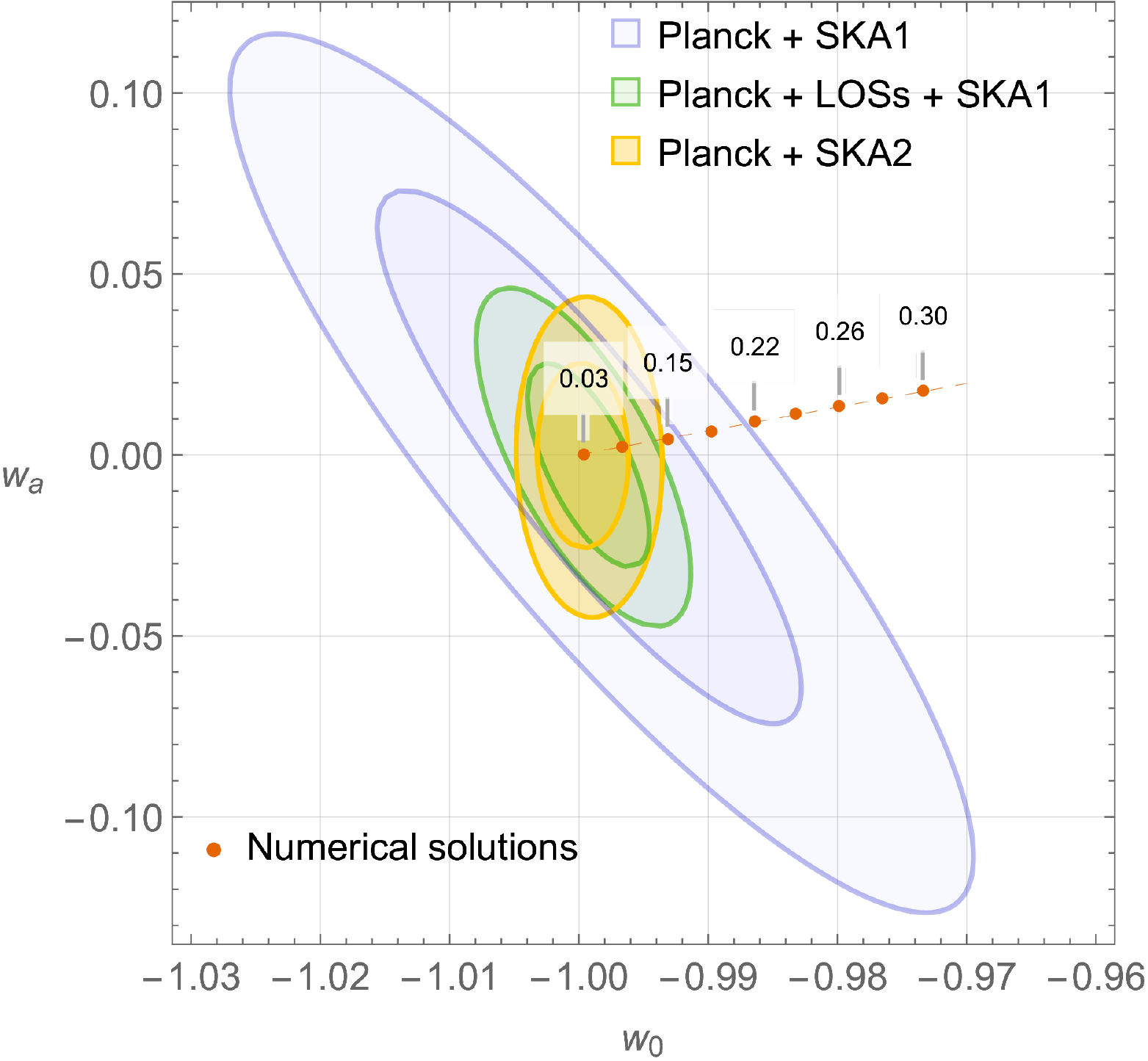}
    \caption{Comparison between the expected 67 and 95\% CL constraints on $w_0$-$w_a$ for various future experiments (blue, green and yellow ellipses) together with numerical results for $\abs{\lambda}$ for tracking freezing models with potentials given by Eq. \eqref{eq:potential_tracking}. The parameter $\alpha$ ($\Gamma=VV''/(V')^2=1+1/\alpha$) is ranging from $0.001$ to $0.08$ and the labels of the points are the values of $\abs*{\lambda}$.}
    \label{fig:tracking}
\end{figure}

From the initial conditions given by Eq. \eqref{eq:IC_tracking}, for the tracker solution, together with the requirements $\Omega_{\phi, 0}\approx 0.7$, numerical solutions can be computed for different values of the free parameter $\alpha$. The results are displayed in Fig.~\ref{fig:tracking}. Numerical results for the $\lambda$ parameter correspond to $\alpha\in [0.001,0.08]$. Since the EOS parameter $w(N)\approx -2/(2+\alpha)$ is nearly constant and began to unfreeze quite recently, numerical simulations for $\alpha > 0.08$ are unnecessary as they fall outside the 95\% CL constraint on $(w_0,w_a)$. One can observe this behavior in Fig.~\ref{fig:stability_tracker}, where $\alpha=1$ was chosen. As previously, the numerical solutions -- that is the points in the $(w_0,w_a)$ plane associated with different $\lambda$ -- are aligned. The upper bounds on $\abs{\lambda}$ are summarized in table \ref{tab:tracking}. In this case also, they are very relevant for the swampland program. 

\begin{table}
\renewcommand{\arraystretch}{1.3}
\begin{tabular}{|c||c|c|c|}
\hline
     & Pl. + SKA1 & Pl. + LOSs + SKA1 & Pl. + SKA2 \\ \hline\hline
67\% CL & $\abs*{\lambda}<0.16$    & $\abs*{\lambda}<0.11$             & $\abs*{\lambda}<0.11$    \\ \hline
95\% CL & $\abs*{\lambda}<0.21$    & $\abs*{\lambda}<0.14$             & $\abs*{\lambda}<0.15$    \\
\hline
%
\end{tabular}
\caption{Expected constraints on $\abs*{V'}/V$ for {\it tracking freezing models}, based on the potential given by Eq. \eqref{eq:potential_tracking}.}
\label{tab:tracking}
\renewcommand{\arraystretch}{1}
\end{table}

\subsection{Scaling freezing models}

\begin{figure}
    \centering
    \includegraphics[width=0.8\linewidth]{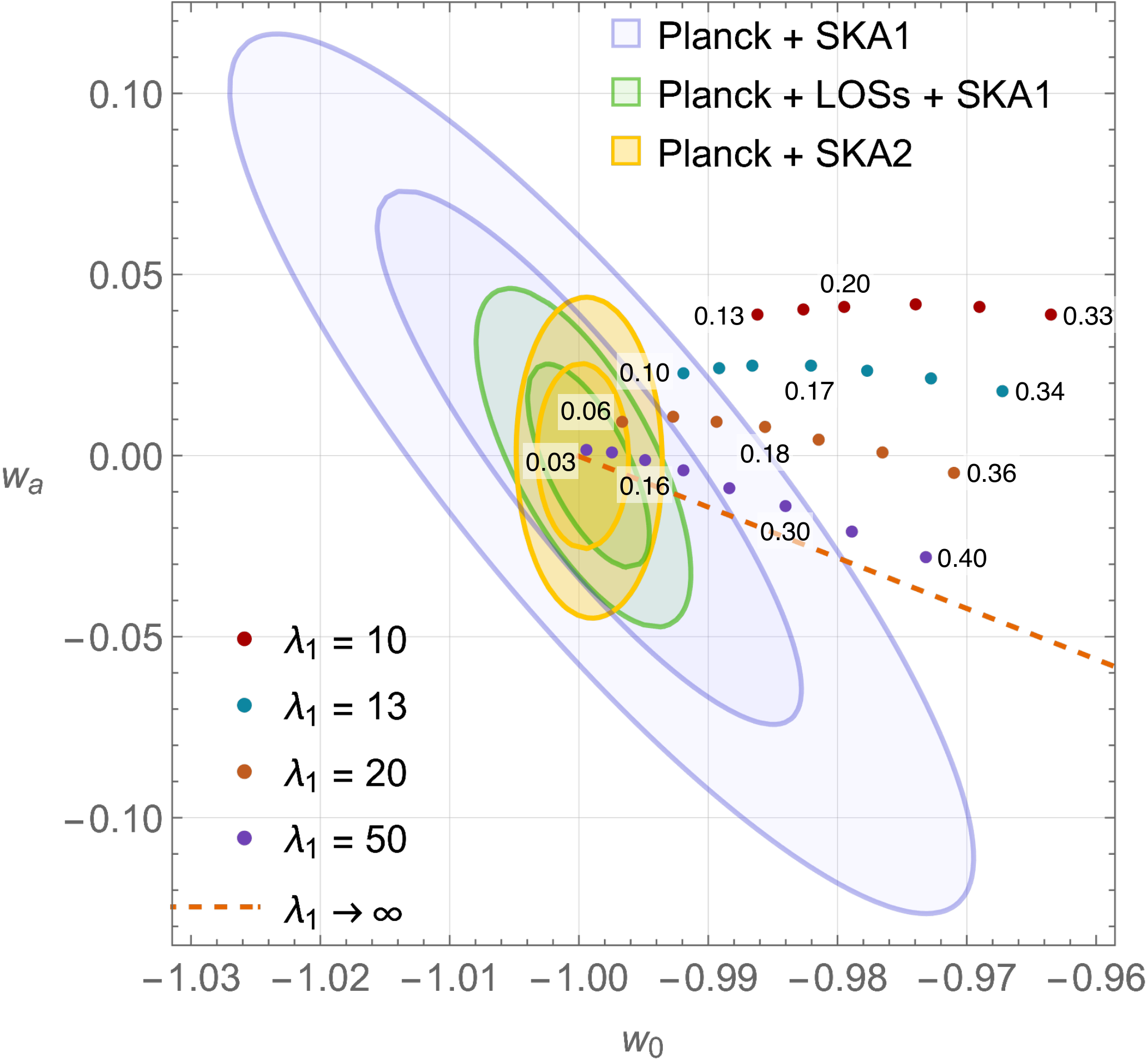}
    \caption{Comparison between the expected 67 and 95\% CL constraints on $w_0$-$w_a$ from various future experiments (blue, green and yellow ellipses) together with numerical results for $\abs*{\lambda}$ for {\it scaling freezing} models with potentials given by Eq.  \eqref{eq:potential_double_exponential}. The parameters $(\lambda_1,\lambda_2)$ are respectively $(10,[0,0.25])$, $(13,[0,0.3])$,  $(20,[0,0.35])$ and $(50,[0,0.4])$ and the labels of the points are the values of $\abs*{\lambda}$.}
    \label{fig:scaling}
\end{figure}

In the case of scaling freezing model, the potential given by Eq. \eqref{eq:potential_double_exponential} has three independent parameters $\lambda_1$, $\lambda_2$ and $V_1/V_2$, making the exhaustive scan {\it a priori} subtler.  However, the value of the ratio between $V_1$ and $V_2$ has nearly no effect on the behavior of the system, hence one can safely set $V_1=V_2$ and $\lambda_1>\lambda_2$ without loss of generality. It was shown in \cite{Barreiro:1999zs,Gupta:2011ip} that in order to have an asymptotic freezing solution with a transition from $w(N\rightarrow -\infty)\approx 0$ to $w(N\rightarrow \infty)\approx -1+\lambda_2^2/3$, it is necessary that $\lambda_1^2>3$ and $\lambda_2^2<3$. The problem therefore basically consists in exploring the cosmological dynamics, and the subsequent minimum of $\abs{V'}/V$, along each track and for values of $(\lambda_1,\lambda_2)$ satisfying the previous constraints. The results for the parameters $(10,[0,0.25])$, $(13,[0,0.3])$, $(16,[0,0.3])$, $(22,[0,0.35])$ and $(50,[0,0.4])$ are displayed in Fig.~\ref{fig:scaling}. Models with $\lambda_1<9.4$ are disfavored from nucleosynthesis analysis \cite{Bean:2001wt} and simulations for  $0.25,0.3,0.35,0.4<\lambda_2<\sqrt{3}$ (respectively, depending on the value of $\lambda_1$) are omitted for clarity as they fall outside the $w_0$-$w_a$ constraints.\\

As expected, the higher the parameter $\lambda_1$, the closer the results to those obtained with the single-exponential potential represented by the dashed line in Fig.~\ref{fig:scaling}. Interestingly, for all finite values of $\lambda_1$, the generic trend is such that the constraints obtained are more stringent that for the single exponential -- that is for a constant $V'$  -- case. To remain conservative, it is therefore possible to keep the results obtained for a single exponential, and given in table \ref{tab:single_exponential}, as generic constraints for {\it scaling freezing} models described by the potential given by \eqref{eq:potential_double_exponential}.

\subsection{Thawing models}
\begin{figure}
    \centering
    \includegraphics[width=0.8\linewidth]{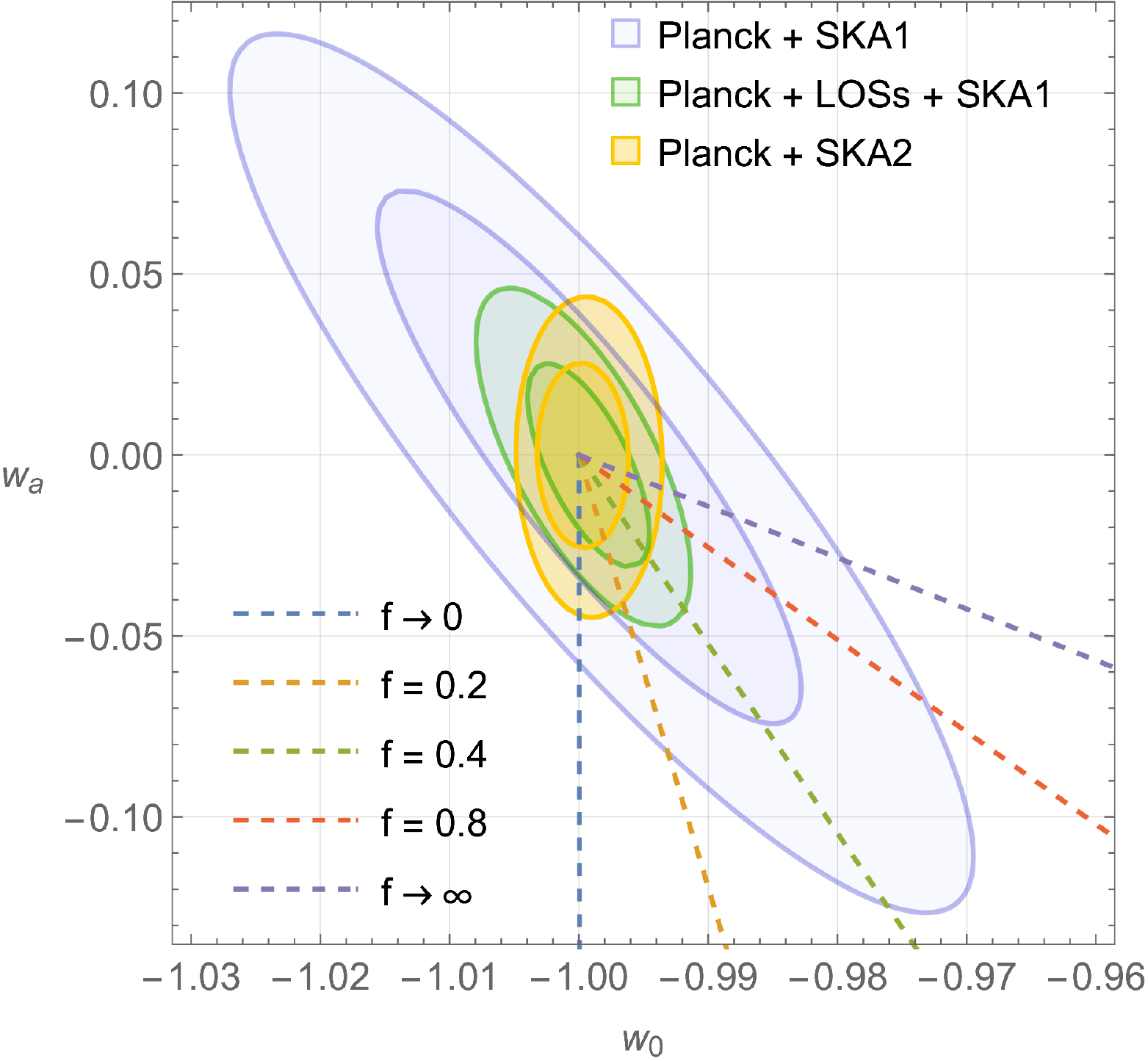}
    \caption{Comparison between the expected 67 and 95\% CL constraints on $w_0$-$w_a$ from various future experiments (blue, green and yellow ellipses) and the $w_0$-$w_a$ relation for {\it thawing model} described by the potential given in Eq. \eqref{eq:potential_cos} for different values of the parameter $f$.}
    \label{fig:thawing1}
\end{figure}

Unlike previous models, for which numerical solutions on $w(N)$ can be fully trusted, thawing models described by potentials given by Eq. \eqref{eq:potential_1pcos} and Eq. \eqref{eq:potential_cos} are better understood using the analytical approximation of Eq. \eqref{eq:approx_w_thawing}. In this approximation, both potentials are equivalent up to the parameter $c_i^2$ and differ by a term $-\lambda_i^2/2$, with $\lambda_i=-V'(\phi_i)/V(\phi_i)$. Starting with the potential given by Eq. \eqref{eq:potential_cos}, where $c^2=c_i^2=1/f^2$ is constant, one can find an explicit relation between $w_0$ and $w_a$ for different values of $f$ using the approximation \eqref{eq:approx_w_thawing}. The result is shown in Fig.~\ref{fig:thawing1}. 
The dashed lines represent allowed values of $w_0$ and $w_a$ for different values of the parameter $f$. In particular, this means that by varying the initial conditions of the system any point on the line can be reached. As discussed earlier, the slope of these lines defines the constraint on the contemporary value of $\lambda=-V'/V$: the steeper the slope, the weaker the constraint. This is problematic as Fig.~\ref{fig:thawing1} shows that  choosing small enough values of $f$ will lead to a very weak constraint on the de Sitter conjecture. However, in this case, the function $\abs{\lambda}$ is increasing with time. This suggests to focus on the remote past so as to set relevant constraints. It might appear as meaningless as the quintessence scalar field was then sub-dominant and the behavior of the Universe was not yet de Sitter-like. If correct, the de Sitter conjecture must however hold at {\it all points} in the field space, as long as the effective field theory description is valid. Interesting constraints on the swampland conjecture can therefore be obtained using contemporary observation of $(w_0,w_a)$ but taking into account the past behavior of the function $\lambda$.\\

\begin{figure}
    \centering
    \includegraphics[width=0.8\linewidth]{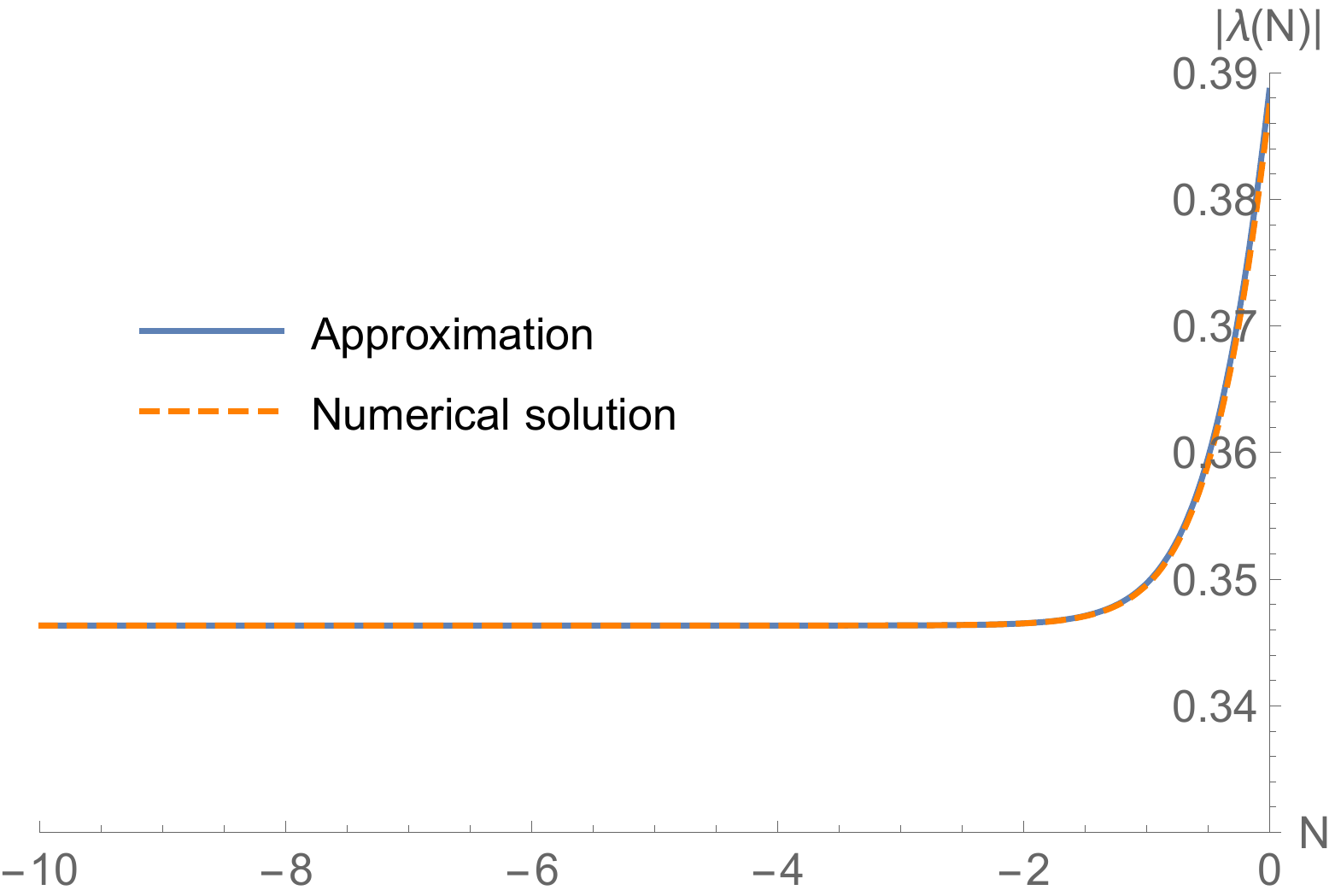}
    \caption{Comparison between the analytical approximation and the numerical solution of $\lambda(N)$ for the thawing model given by Eq. \eqref{eq:potential_cos} with $f=2$. The value $\lambda(0)$ is found using the expected 95\% CL constraint on $w_0$-$w_a$ from the Planck+SKA1 experiment.}
    \label{fig:lambda_thawing}
\end{figure}

Let us call $\lambda_0(f)$ the current value of $-V'/V$ for a given $f$ parameter. A bound can be derived from the intersection between the dashed lines and the ellipses in Fig.~\ref{fig:thawing1} together with Eq. \eqref{eq:w}. Using the approximations of Eqs. \eqref{eq:approx_w_thawing} and \eqref{eq:de-sitter_omega}, as well as the ODE given by Eq. \eqref{eq:l}, one can calculate the behavior of $\lambda$ with respect to the time $N$ (expressed as e-folds) for any parameter $f$. In Fig.~\ref{fig:lambda_thawing}, as an example, $\lambda(N)$ is plotted with the approximation mentioned above, for $f=1$, considering the constraint on $\lambda_0(1)$ from the 95\% CL based Planck+SKA1 observations. The numerical solution is superimposed.
An interesting feature that can be observed on this figure is the stability of $-V'/V$ in the past. The parameter $\lambda(N)$ is indeed quasi-constant for $N<-2$ and one can obtain the constraint of the de Sitter parameter for $f=2$, that is $\abs{V'}/V<0.34$. It is also possible to check in the figure that the approximation of $\lambda(N)$ is strongly reliable.\\

\begin{figure}
    \centering
    \includegraphics[width=0.8\linewidth]{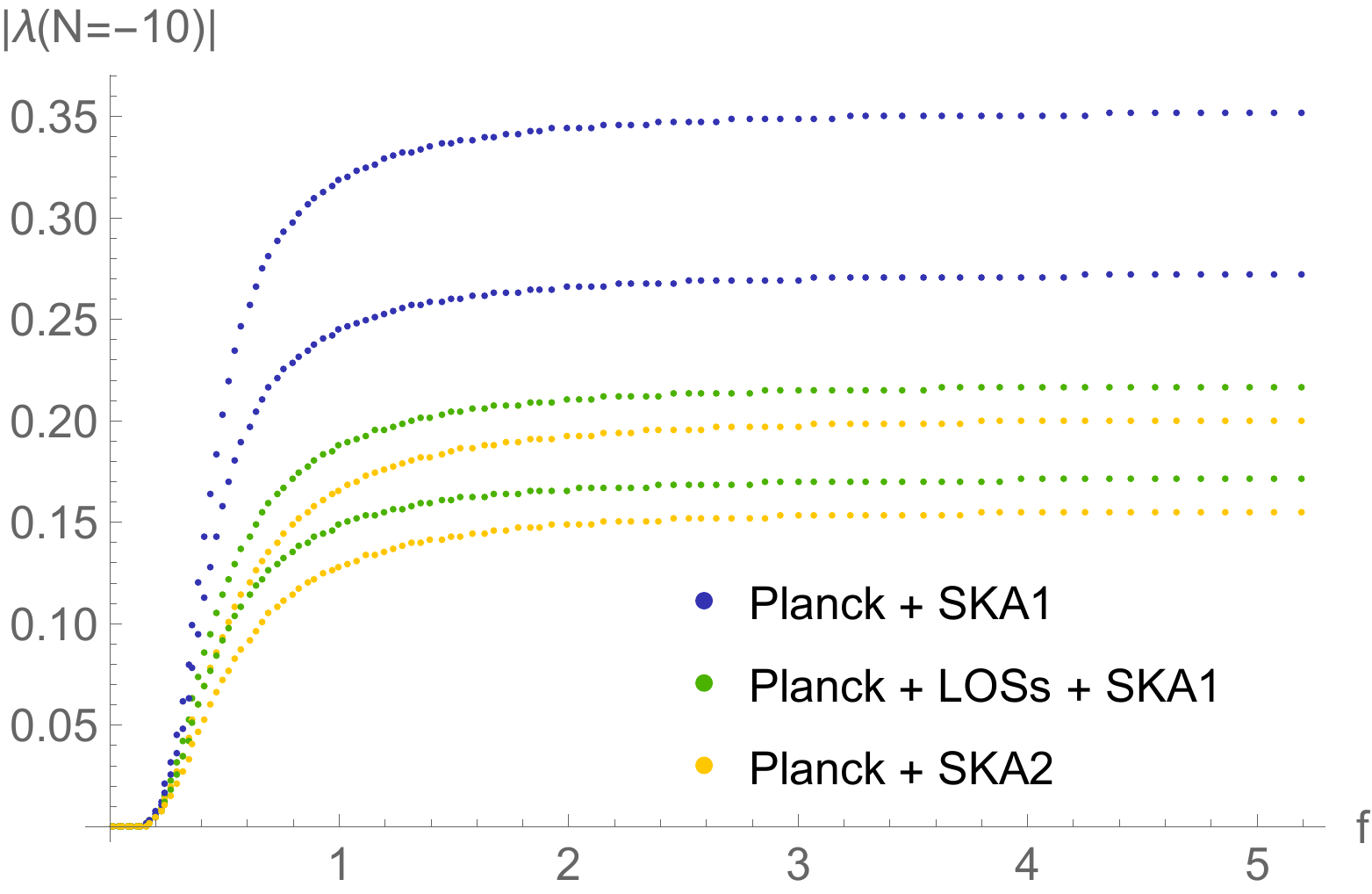}
    \caption{Constraints from various future experiments on $\abs{\lambda(N=-10)}$ as a function of $f$ for {\it thawing} models given by Eq. \eqref{eq:potential_cos}.}
    \label{fig:thawing2}
\end{figure}

This shows the path to the derivation of a constraint actually {\it independent} of the parameter $f$. We first find the current constraint on $-V'/V$, {\it i.e.} $\lambda_0(f)$, we then find the behavior of $\lambda(N)$ and fix $N=-10$ to obtain the constraint on the swampland conjecture. We have checked that considering earlier times does not improve the results. The procedure is repeated for different values of $f$ and for the different experimental scenarios. The results are shown in Fig.~\ref{fig:thawing2} where the limit on $\abs{\lambda(N=-10)}$ is given with respect to $f$. The numerical values are given in table \ref{tab:thawking}. In this case also, they are very stringent and meaningful for the swampland program.\\

\begin{table}
\renewcommand{\arraystretch}{1.3}
\begin{tabular}{|c||c|c|c|}
\hline
     & Pl. + SKA1 & Pl. + LOSs + SKA1 & Pl. + SKA2 \\ \hline\hline
67\% CL & $\abs*{\lambda}<0.27$    & $\abs*{\lambda}<0.17$             & $\abs*{\lambda}<0.16$    \\ \hline
95\% CL & $\abs*{\lambda}<0.35$    & $\abs*{\lambda}<0.22$             & $\abs*{\lambda}<0.20$    \\
 \hline
\end{tabular}
\caption{Expected constraints on $\abs*{V'}/V$ from different sets of experiments for {\it thawing} models with the potential given by Eq. \eqref{eq:potential_cos}. Unlike previous models the limit is obtained in the past. To remain conservative, the limit $f\to \infty$ was considered, which corresponds to the less stringent case.}
\label{tab:thawking}
\renewcommand{\arraystretch}{1}
\end{table}

\subsection{Summary}

This establishes that whatever the (reasonable) potential considered, the future generation of experiments should be able -- if the actual behavior of the Universe is as driven by a cosmological constant -- to put string theory under pressure. At least, the sensitivity is such that the original de Sitter conjecture will be tested in the interesting regime where the measured value is in strong conflict with the theoretical limit. We have shown that for all the classes of models considered, the SKA2 observations are expected to lead to $\abs{V'}/V<0.16$ at 67\% CL and to $\abs{V'}/V<0.20$ at 95\% CL, whereas the de Sitter conjecture requires $\abs{V'}/V\gtrsim 1$.\\

We have also checked that varying the value of $\Omega_{\phi, 0}$ within the observational uncertainties does not change the constraints at the level of accuracy of this work. To summarize, all the considered scenarii for future experiments will contradict the original de Sitter conjecture at a quite high confidence level (unless, of course, the actual dynamics of the Universe reveales not to be driven by a true cosmological constant). \\

If we consider the refined de Sitter conjecture, results are unchanged for {\it tracking freezing} and {\it scaling freezing} models. They always fail to satisfy the new condition. However, in the case of {\it thawing} models, the system is satisfied and no constraint can be put if both conditions are taken into account: when one condition is violated, the other is satisfied. This is an important issue that should be addressed in the future. However, particle physics arguments are expected to allow, in the future, to constrain the $f$ parameter of the potential, therefore breaking the degeneracy \cite{Marsh_2016}. 

\section{Prospects}
Although this work is devoted to the actual estimate of the Vera Rubin, Euclid, and SKA capabilities to improve constraints on the de Sitter conjecture, it is also worth trying to go beyond those experiments. We therefore provide estimates of the upper bound on $\abs{\lambda}$ that could be derived from hypothetical even larger observatories to be possibly constructed in the long run. This also allows one to get an accurate limit on $\abs{\lambda}$ for contours in the $w_0-w_a$ plane that might differ from the simulations used in this study. We, however, still assume an elliptic approximation for the confidence level isolines and denote, respectively, $\sigma_{w_a}$ and $\sigma_{w_0}$ the uncertainties on the semi-major and semi-minor axes, which are considered aligned with the $w_0$-$w_a$ axes, similarly to the Planck+SKA2 constraint. Obviously, other hypotheses could be made but this allows to capture the main features.\\

\begin{figure}
    \centering
    \includegraphics[width=0.8\linewidth]{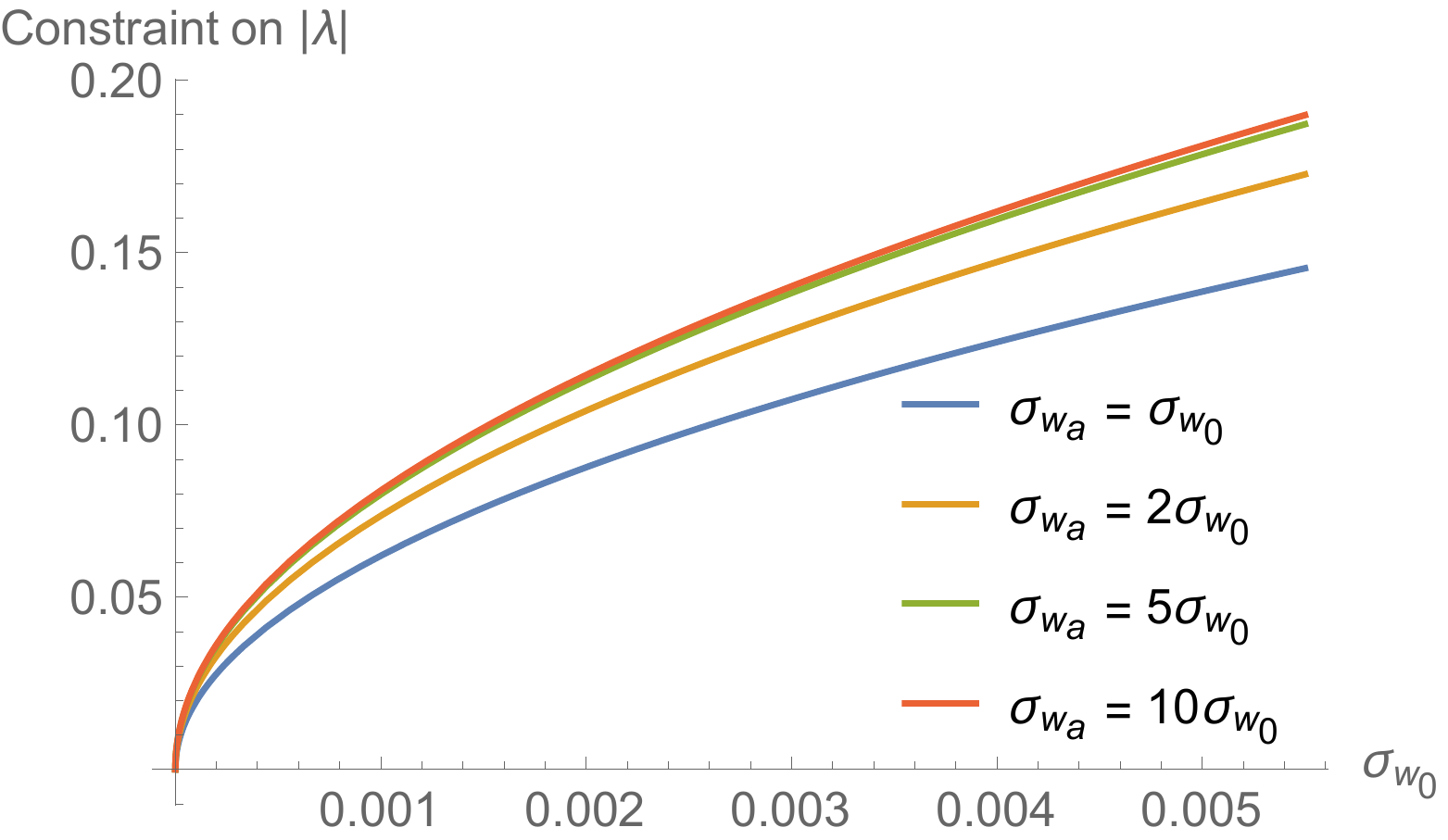}
    \caption{Evolution of the constraint on the de Sitter conjecture with respect to the standard deviation on $w_0$. The standard deviation on $w_a$ is set such that the shape of the ellipse is unchanged.}
    \label{fig:sigma_a_wrt_sigma_0}
\end{figure}

\begin{figure}
    \centering
    \includegraphics[width=0.8\linewidth]{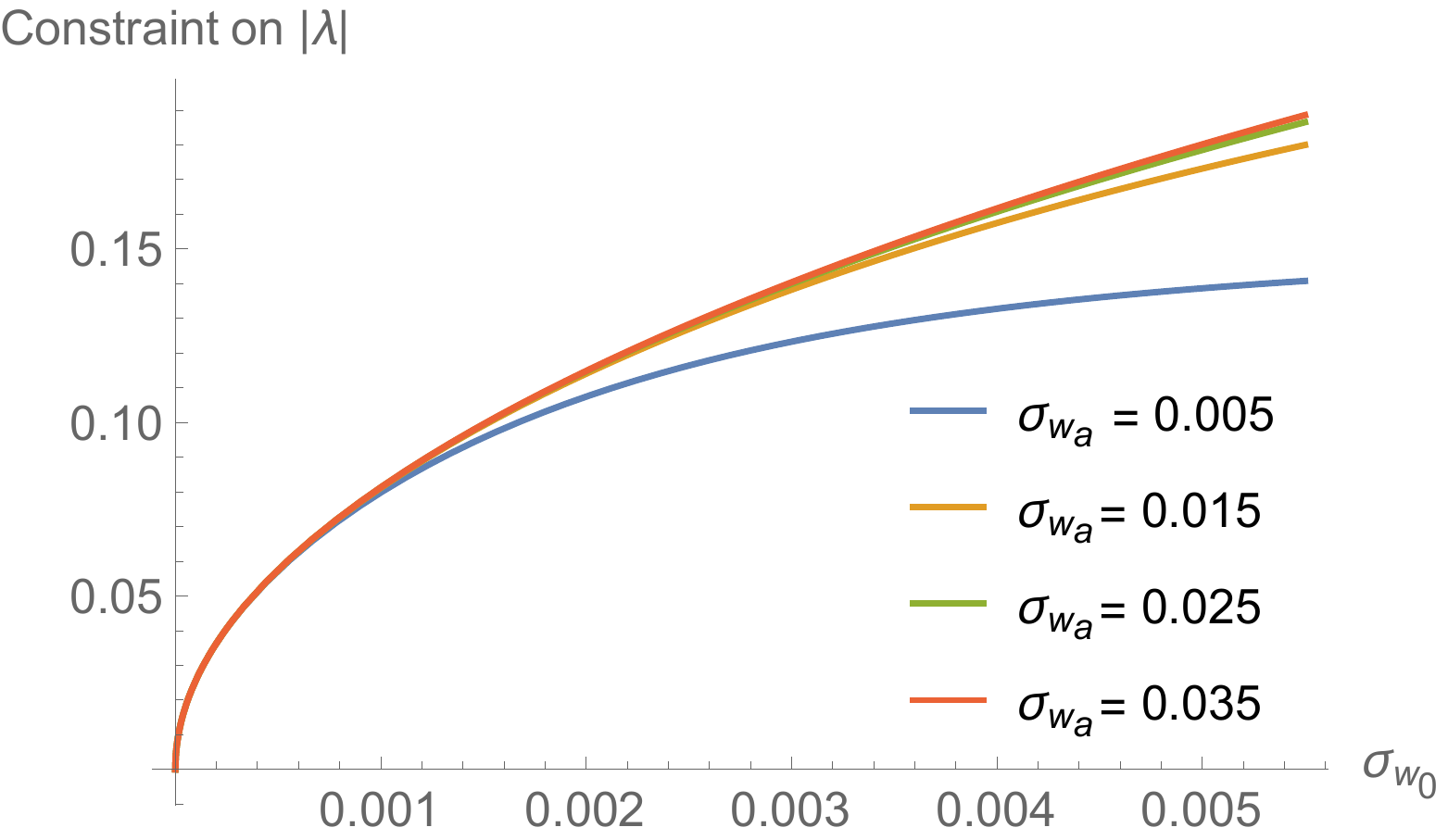}
    \caption{Evolution of the constraint on the de Sitter conjecture with respect to the standard deviation on $w_0$ for different values of the standard deviation on $w_a$.}
    \label{fig:sigma_a_constant}
\end{figure}

\begin{figure}
    \centering
    \includegraphics[width=0.8\linewidth]{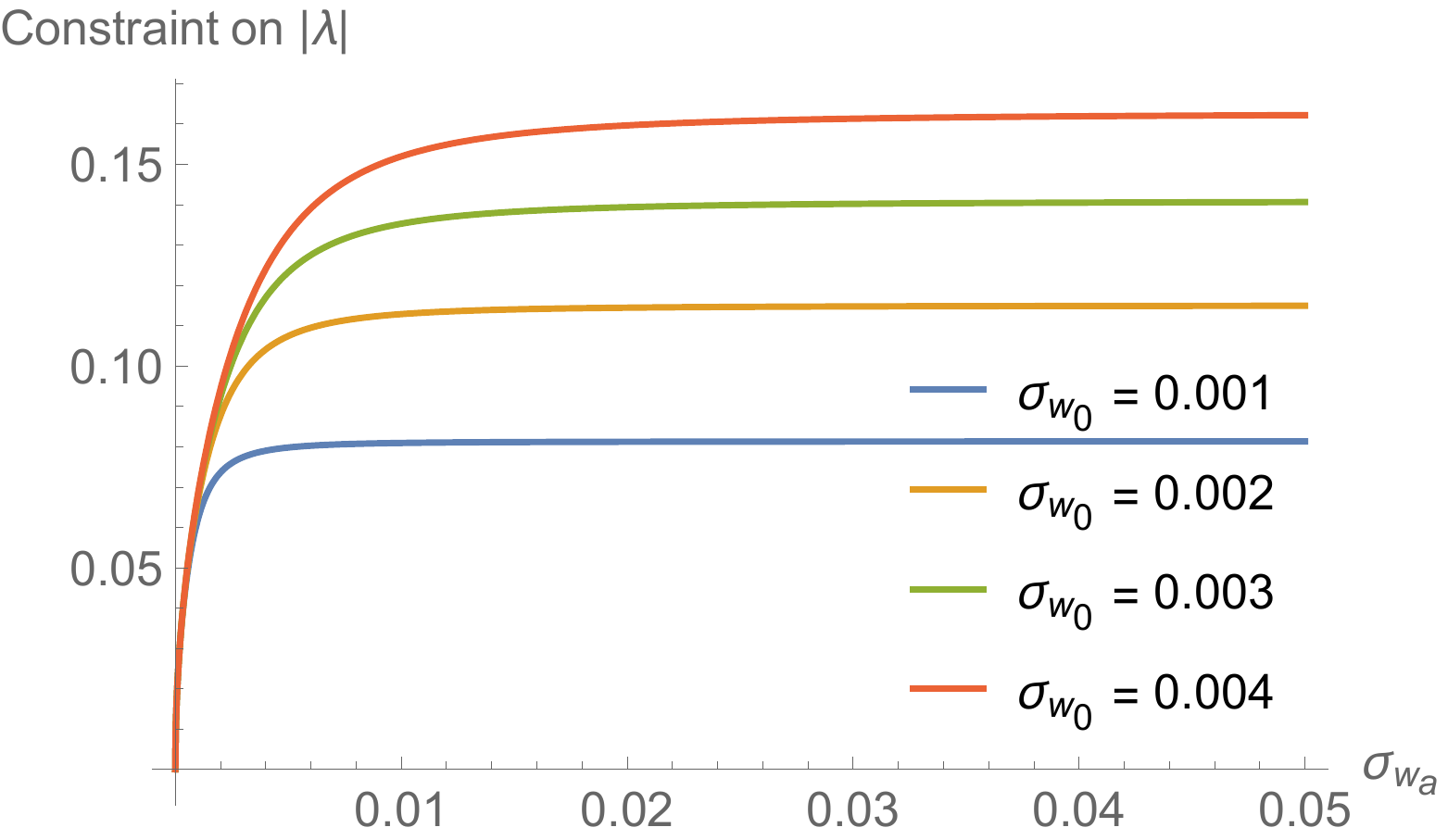}
    \caption{Evolution of the constraint on the de Sitter conjecture with respect to the standard deviation on $w_a$ for different values of the standard deviation on $w_0$.}
    \label{fig:sigma_0_constant}
\end{figure}

In Fig. \ref{fig:sigma_a_wrt_sigma_0}, we display the 67\% CL limit on $\abs{\lambda}$ keeping the shape of the ellipse, {\it i.e.} the ratio between $\sigma_{w_a}$ and $\sigma_{w_0}$, unchanged. For comparison, this ratio is ranging between $7.2$ and $9.3$ for the various expected constraints from Planck+SKA1, Planck+LOSs+SKA1 and Planck+SKA2 simulations. In Fig. \ref{fig:sigma_a_constant}, we display the limit as a function of $\sigma_{w_0}$ for different values of $\sigma_{w_a}$. The other way round, in Fig. \ref{fig:sigma_0_constant}, we plot the limit as a function of $\sigma_{w_a}$ for different values of $\sigma_{w_0}$. We always consider the less constraining potential. 

This shows that there is still room for improving the limits beyond the next generation of experiments. This also underlines that ameliorating the sensitivity on $w_0$ or $w_a$ changes the situation differently. Although reducing the error on $w_0$ has a monotonic and quite regular effect on the improvement on the limit, the $\sigma_{w_a}$ behavior exhibits a ``plateau" as soon as $\sigma_{w_a}>0.01$.

\section{Remarks and conclusion}

\subsection{Is the de Sitter conjecture reliable ?}

The de Sitter conjecture is just a conjecture. How reliable is it (see the introduction of \cite{Raveri:2018ddi}) ? It is known that the landscape of string theory does contain Minkowski solutions. The richness of the structure is nearly infinite. Each geometry can support more than $10^{10^5}$ flux vacua \cite{Taylor:2015xtz} and the number of compactification geometries is higher than $10^{1000}$ \cite{Taylor:2017yqr}. However, constructing metastable de Sitter solutions from this huge landscape still appears as highly problematic and this is the main underlying motivation for the conjecture. It is indeed known that de Sitter space is excluded as a solution of fundamental supergravity theories \cite{Maldacena:2000mw}. It is also neither a solution of type I/heterotic supergravity \cite{Green:2011cn,Gautason:2012tb}, nor of heterotic world-sheet conformal field theory \cite{Kutasov:2015eba}. Basically, it seems quite well established that de Sitter solutions cannot be found in regions of parametric control in string theory \cite{Dine:1985kv,Dine:1985he}. Interesting attempts to evade this conclusion do exist in type IIB string theory \cite{Giddings:2001yu,Andriot:2020wpp}, together as in type IIA and M-theory. None of them is however conclusive either because of the lack of control on the quantum corrections to the effective space-time action (they neglect the fact that the background is non-static \cite{Sethi:2017phn}) or because of the absence of fully explicit constructions.\\

Although anti-de Sitter vacua are well understood in string theory, it is a fact that de Sitter space is surprisingly hard to control. Many no-go theorems  (see, {\it e.g.}, \cite{HariDass:2002si,Russo:2019fnk,Shukla:2019dqd,Basile:2020mpt}) were derived and quite a lot of concrete examples support the possibility that string theory just cannot admit any de Sitter vacuum. Still, counterarguments are being built, around the Kachru-Kallosh-Linde-Trivedi (KKLT) proposal \cite{Kachru:2003aw,Kallosh:2019axr} using a K\"ahler moduli stabilisation or with Large Volume Scenarios (see \cite{Cicoli:2008va} and references therein). The full picture is still unclear.\\

The de Sitter swampland program is an active and highly controversial field of research \cite{Andriot:2018ept}. It might very well be that there is enough complexity in the space of string theory vacua and sufficient richness in the unexplored sectors of the model for a landscape of de Sitter solutions to indeed exist. The extraordinary difficulty to exhibit any convincing de Sitter solution, in spite of the huge number of explored solutions, however suggests that de Sitter space lies in the swampland, and that the conjecture holds. Recently, interesting links with the transplanckian censorship were even built \cite{Bedroya:2019snp}. This is not a theorem but, in our opinion, a reasonable guess. The conjecture constitutes, at least, an outstanding way to possibly put string theory under pressure, something that has proven to be extremely difficult in the last decades.

\subsection{Other approaches to quantum gravity}

In most studies devoted to the swampland (see, {\it e.g.}, \cite{Palti:2019pca}), including this article, the words ``string theory" and ``quantum gravity" are used as if  they were synonymous or, at least, as if quantum gravity was to be understood as a sector of string theory only. This is obviously exagerated. There are many other roads toward quantum gravity  (see \cite{Oriti:2009zz} for an overview): loop quantum gravity (see,  {\it e.g.}, \cite{lqg3}) non-commutative geometry (see, {\it e.g.},  \cite{Chamseddine:1996zu}), group field theory  (see,  {\it e.g.}, \cite{Baratin:2011hp}), causal sets (see, {\it e.g.} \cite{Bombelli:1987aa,Sorkin:2009bp}), asymptotic safety (see,  {\it e.g.}, \cite{Weinberg:2009bg,Saueressig:2019fmx}), causal dynamical triangulation (see,  {\it e.g.}, \cite{Loll:2019rdj}), etc. Those models are unquestionably speculative -- the semiclassical limit is often not known or not clear -- but all have preliminary predictions for cosmology \cite{Barrau:2017tcd}. Most of them have no problem with de Sitter spaces. The existence of a small and positive cosmological constant was even predicted, before it was observed, by Sorkin within the causal sets framework (see references in \cite{Dowker:2017zqj}). It could also be that gravity does not need to be quantized \cite{Tilloy:2019hxe}. 

It should therefore be made clear that the de Sitter conjecture is not about any theory of quantum gravity but only about string theory. This is why we believe that it makes sense, as we did here, to evaluate whether this can be useful to potentially falsify string theory in the future. It is however much more hazardous, as sometimes advertised, to use swampland conjectures to rule out some low-energy models. The fact that they lie in the swampland of string theory does not mean they are intrinsically wrong\footnote{Some swampland arguments are very generic and based on solid quantum field theory conclusions about the UV completion. They have to be distinguished from string-inspired guesses.}: string theory is far from being established. Although intensively debated, the claimed {\it non-empirical corroboration} of string theory is not sufficient to make it a universal paradigm \cite{Dawid:2017rzq,Chall:2018mfn,Cabrera:2018idz}. It seems to us much more fruitful to use the swampland ideas as a first -- and very welcome -- step toward a possible proof of the effective falsifiability of string theory. This is however not even obvious: from the point of view of string theory it is not clear that one should expect dark energy to be described by a scalar field. 

\subsection{Conclusion and future developments}

In this work, we have considered three classes of potentials as benchmarks for quintessence models. We have shown that the SKA network of radio-antennas, the Vera Rubin ground based telescope, and the Euclid satellite could be able to derive a limit on the first de Sitter conjecture, $\abs{V'}/V<0.16$, that contradicts the most reliable theoretical estimates \cite{Andriot:2020lea}. This shows that if the swampland conjecture holds, string theory might be on the road of ``falsification" in the next decade. This conclusion would require the inclusion of multi-field scenarios \cite{Achucarro:2018vey,Bravo:2019xdo,Lin:2021ciy}.

Our main result is nearly model independent in the sense that the upper bound mentioned above is the less stringent one among the three classes of potential considered here. {\it Scaling tracking}, {\it scaling freezing}, and {\it thawing} models are the main ideas currently on the table for quintessence. We have therefore scanned the panel of intensively discussed models. It is however not impossible that other potentials, in agreement with the cosmological dynamics but less constrained from the viewpoint of the de Sitter conjecture, might be found. Our result is therefore not a theorem but a reasonable conclusion based on consistent potentials. It should however be mentioned that, in principle, the actual potential could be reconstructed by a combined cosmological analysis \cite{Boisseau:2000pr}. This opens the possibility to derive a precise measurement of  $-V'/V$ and $V''/V$.

We have also investigated possible long run constraints as a function of the shape and size of the ellipse of uncertainty, beyond the main surveys considered in most of this work.

Showing that the real world does not fulfill the de Sitter conjecture would unquestionably not be enough to discard string theory. But, among other indications, it might play a role in a possible paradigm shift. Low-scale supersymmetry as a fully natural solution to the hierarchy problem was not abandoned by part of the community just after one unsuccessful run at the LHC but after many arguments came weakening the edifice. Every sign counts. 

Obviously, it could also be that string theory is correct, that the de Sitter conjecture holds and that our Universe does not conflict with it but follows a non purely de Sitter dynamics. We leave for a future study the investigation of the detection capability ({\it i.e.} a measurement $|V'|/V\sim\lambda_D$ with $\lambda_D$ not too small) of future surveys at the level required by the conjecture. Is it still possible to measure a ``nearly but not exactly" de Sitter behavior marginally compatible with the Swampland criteria ? The interval of possible values $\lambda_D$, not conflicting with any known data, is quite narrow. It would also be important to take into account the Hubble constant tension in this framework \cite{Banerjee:2020xcn}. 

Finally, on the purely theoretical side, not only should the very validity of the conjecture be better understood but the actual value $\lambda_c$ would have to be better evaluated. The latest estimates are encouraging but one might also argue that our limit of 0.16 is not that far from the expected value of order one. Precise numbers from the string side are now needed beyond orders of magnitude.

\appendix
\section{Appendix : intuitive remarks}

In this section, we summarize some qualitative arguments, to guide the unfamiliar reader, on the reasons why de Sitter spaces are so problematic in string theory. There are no straightforward and fully intuitive explanations. Rather, there are many ``indications" that build up together. The following list (borrowing from \cite{Akrami:2018ylq,Raveri:2018ddi,Heisenberg:2020ywd}) does not pretend to be exhaustive and the arguments do not try to be rigorous.

\begin{itemize}
\item As we have explained before, supersymmetry and de Sitter spaces are not easy to reconcile. Exact supersymmetry is incompatible with the de Sitter symmetries. Finding solutions with spontaneously broken supersymmetry is the natural way to go. In principle, it is possible to start with a theory which explicitly breaks supersymmetry but then one generically encounters stability and divergence issues.
\item In the string framework, the AdS/CFT correspondence plays a central role in paths toward quantum gravity as it states that a strongly-coupled $n$-dimensional gauge theory is equivalent to a gravitational theory in a $(n+1)$-dimensional anti-de Sitter spacetime. It, however, cannot be (at least not in a known and controlled way) extended to de Sitter spacetimes. 
\item Stability is not ensured. If a de Sitter solution is defined as a positive extremum in the fields space, it is not clear that it can correspond to a minimum for {\it all} the fields. This question has been addressed statistically from the viewpoint of ``random supergravity potentials" and ``random matrices" and is not yet fully clear.
\item Some generic arguments suggest that scalar potentials tend to 0 when the string coupling constant (or the volume of internal dimensions) goes to infinity. But if one constructs a de Sitter vacuum -- that is with a positive potential -- while the potential vanishes at infinity, it can only be metastable. The relevant question becomes the one of the lifetime of this vacuum but it is expected that non-perturbative quantum effect will destabilized it.
\item It might be than one of the most ``physical" reason behind the unsuitability of de Sitter spaces in string theory is rooted in the transplanckian censorship conjecture (previously mentioned in the text). Some explicit calculations show that it seems strongly related with the de Sitter conjecture.
\item Finally, whatever the detailed framework within string/M-theory, it becomes clear that the question of constructing de Sitter solutions is extremely constrained. Either relying on the KKLT scenario or on more classical approaches, constraints are accumulating, making constructions incredibly difficult -- if not totally impossible. 
\end{itemize} 

\subsection{Acknowledgements}
We thank David Andriot for enlightening comments on the de Sitter conjecture. We also thank the anonymous referee who helped a lot improving the article.

\bibliography{refs.bib}

 \end{document}